\documentclass[12pt]{article}
\usepackage[centertags]{amsmath}
\usepackage{amsfonts}
\usepackage{amssymb}
\usepackage{amsthm} 
\usepackage{newlfont}
\usepackage{epsfig}
\usepackage{amscd}

\newcommand{\beq}{\begin{equation}}
\newcommand{\eeq}{\end{equation}}
\newcommand{\ba}{\begin{array}}
\newcommand{\ea}{\end{array}}
\newcommand{\bea}{\begin{eqnarray}}
\newcommand{\eea}{\end{eqnarray}}

\begin{document}

\begin{center}
{\large \sc \bf On  integration of multidimensional generalizations of classical $C$- and $S$-integrable nonlinear partial differential equations}

\vskip 15pt

{\large A. I. Zenchuk}

\vskip 8pt

{\it Institute of Chemical Physics, RAS
Acad. Semenov av., 1
Chernogolovka,
Moscow region
142432,
Russia}

\smallskip

\vskip 5pt

e-mail:  { zenchuk@itp.ac.ru }

\vskip 5pt

{\today}

\end{center}

\begin{abstract}
We develop a new integration technique  allowing one to construct a rich manifold of particular solutions to 
multidimensional generalizations of classical $C$- and $S$-integrable Partial Differential Equations (PDEs). Generalizations of (1+1)-dimensional $C$-integrable  and (2+1)-dimensional $S$-integrable $N$-wave equations are derived among examples.
Examples of multidimensional second order PDEs are represented as well.
\end{abstract}

\section{Introduction}

After the original work \cite{GGKM}, the integrability technique was intensively developing. At present time, it covers large class of nonlinear
Partial Differential Equations  (PDEs) applicable in different branches of physics and mathematics. One should mention a few most popular integration methods, such as
$(a)$ the linearization by some direct substitution, for instance, by the Hopf-Cole substitution \cite{HopfCole} and by its multidimensional generalization \cite{Santini} 
(appropriate nonlinear PDEs are referred to as $C$-integrable PDEs, \cite{Calogero1,Calogero2,Calogero3,Calogero4,Calogero5,Calogero6}); $(b)$
 the method of characteristics \cite{Whitham}  and its matrix generalization \cite{SZ,ZS} ($Ch$-integrable PDEs); $(c)$ the inverse spectral transform method  \cite{GGKM,ZMNP,AC}, the dressing method 
 \cite{Konop,ZS1,ZS2,ZM,BM} and Sato approach \cite{OSTT} ($S$-integrable PDEs). $S$-integrable nonlinear PDEs are most applicable in physics. We recall a few types of these equations:  the soliton equations in (1+1)-dimensions, 
such as the Korteweg-de Vries (KdV) \cite{GGKM,KdV} and the Nonlinear Shr\"odinger (NLS) 
\cite{ZS_NLS} equations; the soliton   
(2+1)-dimensional equations, such as 
the Kadomtsev-Petviashvili (KP) \cite{KP} and  the Davey-Stewartson (DS) \cite{DS} equations;
 the self-dual type PDEs having instanton solutions, like the Self-dual Yang-Mills equation (SDYM);
PDEs assotiated with commuting  vector fields \cite{Krichever,TT,DMT,KAR,GMA,MS1,MS2}.
 Nevertheless, the class of completely integrable nonlinear PDEs is very restrictive. Thus, extensions of the integrability technique on new types of nonlinear PDEs is an actual problem.

In this paper we suggest a new version of the dressing method allowing one to construct a rich manifold of particular solutions to new class of nonlinear PDEs in any dimension. The novelty of this class of PDEs is that integrability technique does not generate commuting flows to them  in usual sense, unlike all   methods mentioned above, where any nonlinear PDE appears together with the commuting hierarchy of nonlinear PDEs.
We show that our algorithm may provide   arbitrary functions of $m-1$ independent variables in  the 
 solution space to $m$-dimensional PDEs, which suggests us to consider these PDEs as candidates for complete integrable PDEs. However we do not represent rigorous justification of complete integrability. 

To anticipate, we give simple examples of  nonlinear PDEs for the matrix field $V$, derived in this paper.

1. The system of  first order $D$-dimensional PDEs,
\begin{eqnarray}\label{nl1_intr0}
&&\sum_{m=1}^{D}\left(
V_{t_m} +V C^{(m)}V\right)
B^{(m)}=0,
\end{eqnarray}
where $B^{(m)}$ and $C^{(m)}$ are some constant matrices.
This is multidimensional generalization of $C$-integrable (1+1)-dimensional
 nonlinear $N$-wave equation. 

Simple example of this equation corresponds to $D=2$, $t_1=x$, $t_2=y$, $B^{(1)}_2=B^{(2)}_1=0$,
\begin{eqnarray}\label{intr_V}
V=\left(
\begin{array}{cc}
u&q\cr
p&v
\end{array}
\right).
\end{eqnarray}
 Then eq.(\ref{nl1_intr0}) reads
\begin{eqnarray}\label{intr_ex_1}
&&
u_x+u^2 C^{(1)}_{1} + pq C^{(1)}_2 =0,\\\nonumber
&&
p_x+p(u C^{(1)}_1 + v C^{(1)}_2)=0,\\\nonumber
&&
v_y+v^2 C^{(2)}_{2} + pq C^{(2)}_1 =0,\\\nonumber
&&
q_y+q(u C^{(2)}_1 + v C^{(2)}_2)=0,
\end{eqnarray}
which reduces to the Liouville equation 
\begin{eqnarray}\label{Liouville}
f_{xy}=C^{(1)}_2 C^{(2)}_1\exp(2 f),
\end{eqnarray}
 if $C^{(1)}_1=C^{(2)}_2=0$, $q=p=e^f$. Here and below, $B^{(m)}_\alpha$ and $C^{(m)}_\alpha$ mean the $\alpha$th diagonal elements of the diagonal matrices $B^{(m)}$ and $C^{(m)}$ respectively.

2. The   first order  $D_1 D_2$-dimensional PDEs
\begin{eqnarray}\label{nl1_intr}
&&
\sum_{m_2=1}^{D_2}
\sum_{m_1=1}^{D_1}
L^{(m_1)}\left(
V_{t_{m_1m_2}} +V C^{(m_1m_2)}V\right)
R^{(m_2)}=0,
\end{eqnarray}
where $L^{(m_1)}$, $C^{(m_1m_2)}$ and  $R^{(m_2)}$ are some constant matrices.
 The simple example corresponds to 
$D_1=D_2=2$, 
$L^{(m_1)}_\alpha=R^{(m_2)}_\beta=0$ if $\alpha\neq m_1$ and $\beta\neq m_2$ respectively, $t_{11}=x$, $t_{12}=y$, $t_{21}=z$, $t_{22}=t$.
Let $V$ be given by eq.(\ref{intr_V}), then
eq.(\ref{nl1_intr}) yields:
\begin{eqnarray}\label{intr_ex_2}
&&
u_x+u^2 C^{(11)}_{1} + pq C^{(11)}_2 =0,\\\nonumber
&&
p_z+p(u C^{(21)}_1 + v C^{(21)}_2)=0,\\\nonumber
&&
v_t+v^2 C^{(22)}_{2} + pq C^{(22)}_1 =0,\\\nonumber
&&
q_y+q(u C^{(12)}_1 + v C^{(12)}_2)=0,
\end{eqnarray}
which reduces to the following four-dimensional generalization of Liouville equation (\ref{Liouville})
\begin{eqnarray}\label{Liouvulle_gen}
g_{xy}=C^{(11)}_2 C^{(12)}_1e^{f+g},\;\;\;f_{zt}=C^{(22)}_1 C^{(21)}_2 e^{f+g}
\end{eqnarray}
if  $C^{(11)}_1=C^{(22)}_2=C^{(12)}_2=C^{(21)}_1=0$,
$p=\exp f$, $q=\exp g$.

In particular, eq.(\ref{nl1_intr}) reduces to the following $D_0(D_0-1)/2$-dimensional PDE, $D_1=D_2=D_0$
\begin{eqnarray}\label{S_Q_simple_red_intr}
&&
 \sum_{m_2=1}^{D_0}\sum_{{m_1=1}\atop{m_2>m_1}}^{D_0}\left[i\left( L^{(m_1)} V_{\tau_{m_1m_2}}L^{(m_2)}- 
L^{(m_2)} V_{\tau_{m_1m_2}}L^{(m_1)}\right)+\right.\\\nonumber
&&\left.
 L^{(m_1)}VC^{(m_1m_2)}VL^{(m_2)}  - L^{(m_2)}V C^{(m_1m_2)}  VL^{(m_1)}\right] =0,\\\nonumber
&&
t_{m_1m_2}= -i \tau_{m_1m_2},\;\;V=-V^+,\;\;m_2>m_1,
\end{eqnarray}
where $C^{(m_1m_2)}=-C^{(m_2m_1)}$; $C^{(m_1m_2)}$ and $L^{(m_1)}$   are constant diagonal matrices and $^+$ means hermitian conjugate. This equation has a physical mening describing interaction of $n_0(n_0-1)/2$ waves if $V$ is $n_0\times n_0$ matrix. This is multidimensional generalization of $S$-integrable (2+1)-dimensional
 nonlinear $N$-wave equation.

3. The system of second order $D$-dimensional PDEs,
\begin{eqnarray}\label{nl2_intr1}
&&
\sum_{m,n=1}^{D} \left(
V_{t_nt_m} +(V C^{(n)} V)_{t_m} +VC^{(m)}V_{t_n}+
 VC^{(m)}VC^{(n)}V 
\right) B^{(mn)}=0, 
\end{eqnarray}
where $B^{(mn)}$ are  constant diagonal matrices. 
The scalar version of this equation reads
\begin{eqnarray}\label{nl2_intr1_scalar}
&&
\sum_{m,n=1}^{D} 
V_{t_nt_m}B^{(mn)} +V\sum_{m=1}^{D} V_{t_m}\hat C^{(m)} +
 V^3\hat C 
 =0, \\\nonumber
&&
\hat  C^{(m)} =2\sum_{n=1}^D C^{(n)}B^{(mn)}+\sum_{m=1}^D C^{(m)}B^{(mn)},\;\;
\hat  C =\sum_{m,n=1}^D C^{(m)}C^{(n)}B^{(mn)}.
\end{eqnarray}

Note that not all constant coefficients  may be arbitrary in the above nonlinear PDEs. Constructing particular solutions,  
we will reveal some relations among coefficients. 

Particular examples of nonlinear PDEs, such as eqs. (\ref{intr_ex_1},\ref{intr_ex_2},\ref{S_Q_simple_red_intr},\ref{nl2_intr1_scalar}), will not be considered in this paper. Instead of this, we concentrate on the dressing 
algorithm  allowing one to derive general equations, such as  eqs.(\ref{nl1_intr0},\ref{S_Q_simple_red_intr},\ref{nl2_intr1}), and  study the richness of available solution space for them. 

The structure of this paper is following. We will derive  generalization of the classical $C$-integrable first and second order nonlinear PDEs in Sec.\ref{Section:Cint} with eqs.(\ref{nl1_intr0}) and (\ref{nl2_intr1}) as particular examples. Richness of the solution space to eq.(\ref{nl1_intr0}) will be discussed briefly and 
explicite particular solutions to this equation  with $D=3$ will be given.  Generalization of the classical $S$-integrable nonlinear PDEs 
will be consider in Sec.\ref{Section:Sint} with eqs.(\ref{nl1_intr}) and (\ref{S_Q_simple_red_intr}) as particular examples. Richness of the solution space to eq.(\ref{nl1_intr}) will be discussed briefly and 
explicite particular solutions to this equation  with $D_1=D_2=2$ will be given. Conclusions will be represented in Sec.\ref{Section:Conclusions}.

\section{Generalization of $C$-integrable nonlinear PDEs}
\label{Section:Cint}

\subsection{Starting equations}
\label{Section:general}
Our algorithm is based on the 
following integral equation
\begin{eqnarray}\label{Psi}
&&
P(\mu)*\chi(\mu,\lambda;t)=W(\mu;t)*\chi(\mu,\lambda;t)+W(\lambda;t)\equiv W(\mu;t)*\Big(\chi(\mu,\lambda;t)+{\cal{I}}_1(\mu,\lambda)\Big),
\end{eqnarray}
where
$P(\mu)$, $\chi(\mu,\lambda;t)$, $W(\lambda;t)$ are $n_0\times n_0$ matrix functions of arguments,
 $t_i$ are 
 independent variables of nonlinear PDEs, $t=(t_1,\dots,t_{D})$, ${\mbox{rank}}(P)=n_0$, $\lambda$, $\mu$, $\nu$ are complex parameters.
Here $*$ means the integral operator defined for any  two functions $f(\mu)$ and $g(\mu)$ as follows:
\begin{eqnarray}
\\\label{def_ast}
f(\mu)*g(\mu)=\int f(\mu)g(\mu) d\Omega_1(\mu),
\end{eqnarray}
and $\Omega_1(\mu)$ is some measure. We also introduce unit  ${\cal{I}}_1(\lambda,\mu)$ and inverse $f^{-1}(\lambda,\mu)$ 
operators as follows:
\begin{eqnarray}
&&
f(\lambda,\nu)*{\cal{I}}_1(\nu,\mu) ={\cal{I}}_1(\lambda,\nu)*f(\nu,\mu)=f(\lambda,\mu),\\\nonumber
&&
f(\lambda,\nu) * f^{-1}(\nu,\mu)=f^{-1}(\lambda,\nu) * f(\nu,\mu)={\cal{I}}_1(\lambda,\mu).
\end{eqnarray}
We introduce parameters $t_i$ through the function $\chi(\lambda,\mu;t)$, which is defined as a solution to the following system of linear equations:
\begin{eqnarray}\label{t}
\chi_{t_m}(\lambda,\mu;t)=A^{(m)}(\lambda,\nu)* \chi(\nu,\mu;t)+
\tilde A^{(m)}(\lambda) P(\nu)* \chi(\nu,\mu;t) +A^{(m)}(\lambda,\mu),\;\;m=1,\dots,D,
\end{eqnarray}
where $A^{(m)}(\lambda,\nu)$ and $\tilde A^{(m)}(\lambda)$ are $n_0\times n_0$   matrix functions of arguments.
An important requirement to eq.(\ref{Psi}) is that it must be uniquely solvable with respect to $W(\lambda;t)$,
i.e. operator $*(\chi(\mu,\lambda;t)+{\cal{I}}_1(\mu,\lambda))$ must be invertable. 

Matrices $A^{(m)}$ and $\tilde A^{(m)}$ may not be arbitrary. They have to provide compatibility of the
system (\ref{t}).
We will show that there are two different methods which provide this compatibility. 
The first one (Sec.\ref{Section:classical}) yields classical $C$-integrable nonlinear PDEs, linearizable by the multidimensional version of the Hopf-Cole transformation \cite{HopfCole}, 
while the second method (Sec.\ref{Section:gen}) yields a new type of nonlinear PDEs whose complete integrability is not clarified yet. However, our algorithm supplyes, at least, a rich manifold of particular solutions to these PDEs.

The following theorem is valid for both cases.

{\bf Theorem 2.1}.
Matrix function
$W(\lambda;t)$, obtained as a solution to the integral equation (\ref{Psi}) with $\chi$ defined by eq.(\ref{t})  satisfies the following system of compatible  linear equations
\begin{eqnarray}\label{lin}
&&
E^{(m)}(\lambda;t) :=W_{t_m}(\lambda;t) + V^{(m)}(t) W(\lambda;t) 
+(W(\mu;t)-P(\mu))* A^{(m)}(\mu,\lambda) =0,\\\label{Q_sol}\label{V}
&&
V^{(m)}(t)=(W(\mu;t) -P(\mu))*\tilde A^{(m)}(\mu)
,\;\;
m=1,\dots,D.
\end{eqnarray}

{\bf Proof:} 
To derive eq.(\ref{lin}), we 
differentiate eq.(\ref{Psi}) with respect to $t_m$.  Then, in view of eq.(\ref{t}), 
one gets the following equation:
\begin{eqnarray}\label{lin_1}
E^{(m)}(\mu;t)*(\chi(\mu,\lambda;t)+{\cal{I}}_1(\mu,\lambda)) =0
\end{eqnarray}
where $E^{(m)}$ is defined  in eq.(\ref{lin}).
Since operator $*(\chi(\mu,\lambda;t)+{\cal{I}}_1(\mu,\lambda))$ is invertable, eq.(\ref{lin_1}) yields: $E^{(m)}(\mu;t)=0$, which  coinsides with  eq.(\ref{lin}).
$\blacksquare$

{\it Remark:} Following the classical integrability theory, we refer to eq.(\ref{lin}) as the linear equation for the function $W(\lambda;t)$. However this is not  correct, because functions $V^{(m)}(t)$ are defined in terms of $W(\lambda;t)$ by eq.(\ref{V}). Thus, strictly speaking, eq.(\ref{lin}) is a nonlinear equation for $W(\lambda;t)$.  

System (\ref{lin}) is overdetermined  system of compatible linear equations with potentials $V^{(m)}(t)$ in analogy with the classical integrability theory. In the classical theory, nonlinear PDEs for potentials $V^{(n)}$ may be obtained as compatibility conditions for the  appropriate overdetermined linear system. However, this approach does not work in our case because of the last term in eqs.(\ref{lin}). Instead of this, we suggest a different method of derivation of nonlinear PDEs, see Secs.\ref{Section:Cint_class} and \ref{Section:Cint_new}.

Now we analyze two methods that provide  the compatibility  of system (\ref{t}) and derive nonlinear PDEs assotiated with each of them.

\subsection{First method: classical $C$-integrable nonlinear PDEs}
\label{Section:classical}
\label{Section:Cint_class}
In this subsection we 
write the compatibility condition of eqs.(\ref{t}) as follows: 
\begin{eqnarray}\label{compatA}
\left(A^{(m)}(\lambda,\nu) +\tilde A^{(m)}(\lambda) P(\nu) \right)*\chi_{t_n}(\nu,\mu) =
\left(A^{(n)}(\lambda,\nu) +\tilde A^{(n)}(\lambda) P(\nu) \right)*\chi_{t_m}(\nu,\mu),\;\;\forall \; n,m .
\end{eqnarray}
Substituting eqs.(\ref{t}) for derivatives of $\chi$ into eq.(\ref{compatA}) we obtain the following equation:
\begin{eqnarray}\label{LL}
&&
(L^{(m)}*L^{(n)}-L^{(n)}*L^{(m)})*\chi + L^{(m)}* A^{(n)} -L^{(n)}* A^{(m)} =0,\\\nonumber
&&
L^{(m)}(\lambda,\mu)=A^{(m)}(\lambda,\mu) +\tilde A^{(m)}(\lambda) P(\mu), \;\;.
\end{eqnarray}
Let  eq.(\ref{LL})  be satisfied for any function $\chi(\lambda,\mu;t)$ (which is a solution to the system (\ref{t})). Then eq.(\ref{LL}) is equivalent to 
  two following equations relating matrix functions $A^{(m)}$, $\tilde A^{(m)}$ and $P$:
\begin{eqnarray}\label{LA1}
&&
L^{(m)}* A^{(n)} -L^{(n)} *A^{(m)}=0,\\\label{LA2}
&&
L^{(m)}*L^{(n)}-L^{(n)}*L^{(m)}=0\;\;\;\stackrel{{\mbox{eq.\ref{LA1}}}}{\Rightarrow} \\\nonumber
&&
\Big(L^{(m)}(\lambda,\nu) *\tilde A^{(n)}(\nu)  -
L^{(n)} (\lambda,\nu)*\tilde A^{(m)}(\nu)\Big) P(\mu) =0
\end{eqnarray}
Since ${\mbox{rank}}(P)=n_0$, eq.(\ref{LA2}) is equivalent to the following one:
\begin{eqnarray}\label{LA_AL}
&&
L^{(m)} * \tilde A^{(n)}-L^{(n)}* \tilde A^{(m)} =0\;\;\Rightarrow\\\nonumber
&&
A^{(m)} *\tilde A^{(n)} -A^{(n)}* \tilde A^{(m)} = \tilde A^{(n)} P* \tilde A^{(m)} -
 \tilde A^{(m)} P *\tilde A^{(n)},\;\;
n,m=1,\dots,D.
\end{eqnarray}
Eqs.(\ref{LA1}) and (\ref{LA_AL}) represent two constraints on the functions $A^{(m)}(\lambda,\mu)$ and 
 $\tilde A^{(m)}(\lambda)$.

Now we have everything for derivation of nonlinear PDEs for the fields $V^{(m)}(t)$. For this purpose,  let us 
consider the following combination of eqs.(\ref{lin}): 
\begin{eqnarray}
&&
E^{(m)}(\lambda;t)* \tilde A^{(n)}(\lambda)-E^{(n)}(\lambda;t)* \tilde A^{(m)}(\lambda),
\end{eqnarray}
which yields, in view of eq.(\ref{LA_AL})
\begin{eqnarray}
\label{nlin_cl}
&&
V^{(n)}_{t_m}(t)- V^{(m)}_{t_n}(t)+V^{(m)}(t) V^{(n)}(t)-V^{(n)}(t) V^{(m)}(t) =0.
\end{eqnarray}
This equation is known to be linearizable by the Hopf-Cole transformation \cite{HopfCole,Santini}:
\begin{eqnarray}\label{clPsi}
E^{(n)}_H:=\Psi_{t_n}(t)=  \Psi(t)V^{(n)}(t),
\end{eqnarray}
where $\Psi(t)$ is an arbitrary $n_0\times n_0$ matrix function of all variables $t_i$.
The presence of an arbitrary function $\Psi(t)$ is assotiated with the fact that the system of nonlinear PDEs (\ref{nlin_cl}) is not complete. One needs one more equation relating $V^{(n)}$, $n=1,\dots,D$. 

To  derive this additional equation  we introduce either 
additional relations among $A^{(m)}$ and $\tilde A^{(m)}$ in our algorithm based on eq.(\ref{Psi}) or an
additional linear PDE  for  $\Psi$ in the classical algorithm based on eq.(\ref{clPsi}) \cite{Santini}.

For instance, let
\begin{eqnarray}\label{t_cl_0}
\sum_{m=1}^{D} A^{(m)}(\lambda,\nu)* \tilde A^{(n)}(\nu) B^{(m)} =
-\sum_{m=1}^{D}\tilde A^{(m)}(\lambda) P(\nu)*\tilde A^{(n)}(\nu) B^{(m)}
\end{eqnarray}
and/or
\begin{eqnarray}\label{t_cl_1}
\sum_{m=1}^{D}  \Psi_{t_m} B^{(m)} =0,
\end{eqnarray}
where $B^{(m)}$ are $n_0\times n_0$ arbitrary  constant matrices.
Then both the combination of eqs.(\ref{lin}), $\sum_{m=1}^{D}  (E^{(n)}) B^{(m)}$,
and the appropriate combination of eqs.(\ref{clPsi}), $\Psi^{-1}\sum_{m=1}^{D}  (E^{(n)}_H)_{t_m} B^{(m)}$,
yield the same nonlinear  equation for $V^{(n)}$:
\begin{eqnarray}\label{Nwave_constr}
\sum_{m=1}^{D}  \left( V^{(n)}_{t_m} +  V^{(n)}  V^{(m)}\right)B^{(m)}=0,\;\;n=1,\dots,D. 
\end{eqnarray}
This $N$-wave type equation is supplemented by constraints (\ref{nlin_cl}) \cite{Santini}.

In particular, introducing reduction $V^{(m)}(t)=V(t) C^{(m)}$  (where $C^{(m)}$ are $n_0\times n_0$ constant matrices) we reduce  the system (\ref{nlin_cl}) into the following one:
\begin{eqnarray}
\label{nlin_cl_Nwave}
&&
V_{t_m} C^{(n)}- V_{t_n} C^{(m)}+VC^{(m)} VC^{(n)}-VC^{(n)} VC^{(m)} =0,\;\;\forall\;n,m,
\end{eqnarray}
which is a (1+1)-dimensional hierarchy of commuting $C$-integrable $N$-wave equations.

Higher order nonlinear PDEs may be obtained in a similar way introducing appropriate equations instead of eq.(\ref{t_cl_0}) and/or eq.(\ref{t_cl_1}).

$C$-integrable PDEs will not be considered in this paper.

\subsection{Second method: new class of nonlinear PDEs}
\label{Section:gen}
\label{Section:Cint_new}
In this section we represent another way to provide the compatibility  of eqs.(\ref{t}). As a result we obtain a new class of nonlinear PDEs together with the rich manifold of particular solutions. In particular,   there are  solutions in the form of rational functions of  exponents.

Let us 
use the following representation of $A^{(m)}(\lambda,\mu)$ and $\tilde A^{(m)}(\lambda)$:
\begin{eqnarray}\label{A_alp_bet}\label{S_A_alp_bet}
&&
A^{(m)}(\lambda,\mu)=\alpha^{(m)}(\lambda,\nu)\star \beta^{(m)}(\nu,\mu),\;\;\tilde A^{(m)}(\lambda)=
\alpha^{(m)}(\lambda,\nu) \star \tilde \beta^{(m)}(\nu),
\end{eqnarray}
where $\alpha^{(m)}(\lambda,\nu)$, $\beta^{(m)}(\nu,\mu)$ and  $\tilde \beta^{(m)}(\nu)$  are $n_0\times n_0$ 
 matrix functions of arguments.
Here we introduce one more integral operator $\star$ defined  for any two functions $f(\mu)$ and $ g(\mu)$ as follows: 
\begin{eqnarray}
f(\mu)\star g(\mu) \equiv
\int f(\mu) g(\mu) d\Omega_2(\mu),
\end{eqnarray}
where $\Omega_2(\mu)$ is some measure.
We also introduce unit  ${\cal{I}}_2(\lambda,\mu)$ and inverse $f^{-1}(\lambda,\mu)$ operators as follows:
\begin{eqnarray}
&&
f(\lambda,\nu)\star{\cal{I}}_2(\nu,\mu) ={\cal{I}}_2(\lambda,\nu)\star f(\nu,\mu)=f(\lambda,\mu),\\\nonumber
&&
f(\lambda,\nu) \star f^{-1}(\nu,\mu)=f^{-1}(\lambda,\nu) \star f(\nu,\mu)={\cal{I}}_2(\lambda,\mu).
\end{eqnarray}
Now let us write eq.(\ref{t}) using representation (\ref{A_alp_bet}) in the following form:
\begin{eqnarray}
\label{t_xi}
\chi_{t_m}(\lambda,\mu;t)=\alpha^{(m)}(\lambda,\nu)\star\xi^{(m)}(\nu,\mu;t).
\end{eqnarray}
Here
\begin{eqnarray}
\label{xi_m}
&&
\xi^{(m)}(\lambda,\mu;t)=\left(\beta^{(m)}(\lambda,\nu) +\tilde \beta^{(m)}(\lambda) P(\nu)\right)*\chi(\nu,\mu;t) +\beta^{(m)}(\lambda,\mu) ,\;\;m=1,\dots,D.
\end{eqnarray}
It is obvious that the compatibility condition of the system (\ref{t}) is equivalent to the compatibility condition of the system (\ref{t_xi})   which reads
\begin{eqnarray}\label{alp_xi_t}
\alpha^{(m)}(\lambda,\nu)\star \xi^{(m)}_{t_n}(\nu,\mu;t) = \alpha^{(n)}(\lambda,\nu)\star \xi^{(n)}_{t_m}(\nu,\mu;t) ,\;\;\;\forall \;n,m
\end{eqnarray}
instead of eq.(\ref{compatA}).
To satisfy this condition we assume the following relations  between
$\xi^{(m)}$ and $\xi^{(1)}$:
\begin{eqnarray}\label{alpha_xi1}
&&\xi^{(m)}(\lambda,\mu;t)=\eta^{(m)}(\lambda,\nu)\star \xi^{(1)}(\nu,\mu;t),\;\;m>1,\;\;\eta^{(1)}(\lambda,\mu)={\cal{I}}_2(\lambda,\mu),
\end{eqnarray}
where $\eta^{(m)}(\lambda,\mu)$ are some $n_0\times n_0$ matrix functions, which will be specified below.
We also define $t$-dependence of $\xi^{(1)}$ as follows:
\begin{eqnarray}
\label{alpha_xi2}
&&\xi^{(1)}_{t_m}(\lambda,\mu;t) = T^{(m)}(\lambda,\nu)\star  \xi^{(1)}(\nu,\mu;t),
\end{eqnarray}
where $T^{(m)}(\lambda,\nu)$ are some $n_0\times n_0$ matrix functions.
Substituting eqs.(\ref{alpha_xi1}) and (\ref{alpha_xi2}) into eq.(\ref{alp_xi_t}) we obtain the following representation for $\alpha^{(m)}$, $m>1$:
\begin{eqnarray}
\label{alpha_m}
&&
\alpha^{(m)}(\lambda,\mu)=\alpha^{(1)}(\lambda,\nu)\star T^{(m)}(\nu,\tilde\nu)\star (T^{(1)})^{-1}(\tilde\nu,\tilde\mu) \star (\eta^{(m)})^{-1}(\tilde\mu,\mu).
\end{eqnarray}
In turn, the compatibility condition of eqs.(\ref{alpha_xi2}) requires
\begin{eqnarray}
&&
T^{(m)}\star T^{(n)}-T^{(n)}\star T^{(m)}=0,\;\;\Rightarrow \\\label{com_1}
&&
T^{(m)}(\lambda,\mu)=\Big(T(\lambda,\nu)\tau^{(m)}(\nu)\Big)\star T^{-1}(\nu,\mu),\;\;[\tau^{(m)}(\nu),\tau^{(n)}(\nu)]=0,\;\;\forall\;n,m,
\end{eqnarray}
where  $T(\lambda,\mu)$ and $\tau^{(m)}(\mu)$
  are some  $n_0\times n_0$ matrix functions of arguments. Thus compatibility condition of the system (\ref{t_xi}) generates 
eqs.(\ref{alpha_xi1}-\ref{com_1}).

Now, integrating eq.(\ref{alpha_xi2}) with $m=1$  we obtain:
\begin{eqnarray}
\label{xi_expl}
&&
\xi^{(1)}(\lambda,\mu;t)=\left(T(\lambda,\nu)e^{\sum_{i=1}^D\tau^{(i)}(\nu)t_i}\right)\star T^{-1}(\nu,\tilde\nu)\star\xi_0(\tilde\nu,\mu).
\end{eqnarray}
Finally, 
integrating eq.(\ref{t_xi}) with $m=1$ we derive the following explicite formula for $\chi(\lambda,\mu;t)$:
\begin{eqnarray}\label{chi}
\chi(\lambda,\mu;t)=\alpha^{(1)}(\lambda,\nu)\star (T^{(1)})^{-1}(\nu,\tilde\nu)\star \xi^{(1)}(\tilde\nu,\mu;t)+\chi_0(\lambda,\mu),
\end{eqnarray}
where $\chi_0(\lambda,\mu)$ is  $n_0\times n_0$ matrix integration constant.

It is convenient to rewrite expressions for $A^{(m)}$ and $\tilde A^{(m)}$ (see eqs.(\ref{A_alp_bet})) 
using eq.(\ref{alpha_m}) as follows:
\begin{eqnarray}\label{AGamma}
&&
A^{(m)}(\lambda,\mu)=\alpha^{(1)}(\lambda,\nu)\star \Gamma^{(m)}(\nu,\mu),\;\;\tilde A^{(m)}(\lambda)=
\alpha^{(1)}(\lambda,\nu)\star \tilde \Gamma^{(m)}(\nu),\\\nonumber
&&
\Gamma^{(m)}(\lambda,\mu)= T^{(m)}(\lambda,\nu)\star (T^{(1)})^{-1}(\nu,\tilde\nu) \star(\eta^{(m)})^{-1}(\tilde\mu,\tilde\lambda)
\star \beta^{(m)}(\tilde\lambda,\mu),\;\;\\\nonumber
&&
\tilde \Gamma^{(m)}(\lambda)= T^{(m)}(\lambda,\nu) \star(T^{(1)})^{-1}(\nu,\tilde\nu)\star (\eta^{(m)})^{-1}(\tilde\nu ,\tilde\mu)\star\tilde \beta^{(m)}(\tilde\mu).
\end{eqnarray}
Since $\eta^{(1)}(\lambda,\mu)={\cal{I}}_2(\lambda,\mu)$, one has $\Gamma^{(1)}=\beta^{(1)}$ and 
 $\tilde \Gamma^{(1)}=\tilde \beta^{(1)}$.
Representations (\ref{AGamma}) will be used in the rest of Sec.\ref{Section:Cint_new}.

\subsubsection{Internal constraints for $\Gamma^{(m)}$, $\tilde \Gamma^{(m)}$, $T$ and  $\tau^{(m)}$} 
\label{Section:C_internal}
Remember that  definition of $\xi^{(m)}$ in terms of $\chi$, see eq. (\ref{xi_m}),
must be consistent with eq.(\ref{chi}). This requirement generates a set of constraints on  $\Gamma^{(m)}$, $\tilde \Gamma^{(m)}$, $T$ and  $\tau^{(m)}$. To derive these constraints, we
apply operator $(\beta^{(m)}+\tilde \beta^{(m)} P) *$ to eq.(\ref{chi})  from the left. One gets
\begin{eqnarray}
\xi^{(m)}-\beta^{(m)} =(\beta^{(m)}+\tilde \beta^{(m)} P)*\alpha^{(1)}\star (T^{(1)})^{-1}\star \xi^{(1)} +
(\beta^{(m)}+\tilde \beta^{(m)} P) *\chi_0.
\end{eqnarray} 
Substituting eq.(\ref{alpha_xi1}) for $\xi^{(m)}$ one has to obtain an identity valid for any $\xi^{(1)}$.
This requirement yields, first of all, the following equation relating $\beta^{(m)}$, $\tilde \beta^{(m)}$  and $\chi_0$ (the first constraint on the functions $\Gamma^{(m)}$ and $\tilde \Gamma^{(m)}$):
\begin{eqnarray}
&&
\left(\beta^{(m)}+\tilde \beta^{(m)} P\right)*\chi_0 + \beta^{(m)}=0 \;\;\Rightarrow \\\label{beta_m}
&&
\left(\Gamma^{(m)}(\lambda,\nu)+\tilde \Gamma^{(m)}(\lambda) P(\nu)\right)*\chi_0(\nu,\mu) + \Gamma^{(m)}(\nu,\mu)=0,
\end{eqnarray}
and the following definition of $\eta^{(m)}$:
\begin{eqnarray}\label{eta_m}
&&
\eta^{(m)}=\left(\beta^{(m)}+\tilde \beta^{(m)} P\right)*\alpha^{(1)}\star
(T^{(1)})^{-1},\;\;
m=1,\dots,D.
\end{eqnarray}
Finally, eq.(\ref{eta_m}) written in terms of $\Gamma^{(m)}$ and $\tilde \Gamma^{(m)}$ yields:
\begin{eqnarray}\label{eta_m_Gamma}
&&
T^{(m)}(\lambda,\mu)=T^{(1)}(\lambda,\nu)*\left(\Gamma^{(m)}(\nu,\tilde\nu)+\tilde \Gamma^{(m)}(\nu) P(\tilde\nu)\right)*\alpha^{(1)}(\tilde\nu,\tilde\mu)\star
(T^{(1)})^{-1}(\tilde\mu,\mu),\\\nonumber
&&
m=1,\dots,D,
\end{eqnarray}
where $T^{(m)}$ are represented by eq.(\ref{com_1}) in terms of $T$ and $\tau^{(m)}$.
This is the second constraint imposed on the matrix functions  $\Gamma^{(m)}$, $\tilde \Gamma^{(m)}$, $T$ and $\tau^{(m)}$.

Constraints obtained in this section are produced, generally speaking, by   system (\ref{t}) and its compatibility condition. For this reasong we refer to them  as the internal constraints. In contrust, the external constraints
 will be introduced "by hand" in order to derive the nonlinear PDEs, see Theorems 2.2 and 2.3.

\subsubsection{First order nonlinear PDEs for the functions $V^{(m)}(t)$, $m=1,\dots,D$}
{\bf Theorem 2.2. } In addition to eqs.(\ref{Psi},\ref{t}), we impose the following external constraint for the  matrix functions $A^{(m)}(\lambda,\mu)$ and $\tilde A^{(m)}(\lambda)$:
\begin{eqnarray}\label{constrain1}
\sum_{m=1}^{D}A^{(m)}(\lambda,\nu)* \tilde A^{(n)}(\nu) B^{(mn)} =\sum_{m=1}^{D}
\tilde A^{(m)}(\lambda) P^{(mn)},\;\;n=1,\dots,D,
\end{eqnarray}
where $B^{(mn)}$ and $P^{(mn)}$ are some $n_0\times n_0$ constant matrices.
Then $n_0\times n_0$ matrix functions $V^{(m)}(t)$, $m=1,\dots,D$, are solutions to the following system of nonlinear PDEs:
\begin{eqnarray}\label{Q}
&&
\sum_{m=1}^{D}\left[\left(
V^{(n)}_{t_m} +V^{(m)}V^{(n)}\right)
B^{(mn)}+ V^{(m)} \Big({\cal{A}}^{(n)}B^{(mn)} + P^{(mn)}\Big)\right]=0,\;\;n=1,\dots,D,\\\nonumber
&&{\cal{A}}^{(n)} =
P(\lambda)* \tilde  A^{(n)}(\lambda).
\end{eqnarray}

{\bf Proof:} 
Applying operator $* \tilde A^{(n)}$ to the eq.(\ref{lin})  from the right one gets 
the following equation:
\begin{eqnarray}\label{2U}
&&
E^{(mn)}(t):=V^{(n)}_{t_m}(t) +V^{(m)}(t) \Big(V^{(n)}(t)+{\cal{A}}^{(n)}\Big)+U^{(mn)}(t)
=0,
\end{eqnarray}
which introduces a new set of fields $U^{(mn)}$,
\begin{eqnarray}\label{UW}\label{S_UW}
U^{(mn)}(t) = \Big(W(\mu;t)-P(\mu)\Big)*A^{(m)}(\mu,\lambda)*\tilde A^{(n)}(\lambda).
\end{eqnarray}
Due to the relation (\ref{constrain1}), we may eliminate these fields using a 
proper combinations of eqs.(\ref{2U}). Namely, combination $ \sum_{m=1}^{D} E^{(mn)}(t) B^{(mn)}$ results in the system 
(\ref{Q}). $\blacksquare$

\paragraph{Reduction 1.}
Let
\begin{eqnarray}\label{1st_order_ABP}
 P^{(mn)}
=-{\cal{A}}^n B^{(mn)} .
\end{eqnarray}
Then  eq.(\ref{Q}) reduces as follows:
\begin{eqnarray}\label{Q_simple}
&&
\sum_{m=1}^{D}\left(
V^{(n)}_{t_m} +V^{(m)}V^{(n)}\right)
B^{(mn)}=0
\end{eqnarray}
 Note that eq.(\ref{Q_simple}) coinsides with the linearizable eq.(\ref{Nwave_constr}) if $B^{(mn)}=B^{(m)}$. However, eq.(\ref{Nwave_constr}) is supplemented by constraints (\ref{nlin_cl}), which are not valid for eq.(\ref{Q_simple}) in general. 
Constraint (\ref{constrain1}) reads in this case:
\begin{eqnarray}\label{constrain1_red1}
\sum_{m=1}^{D}(A^{(m)}(\lambda,\nu) +\tilde A^{(m)}(\lambda) P(\nu)) * \tilde A^{(n)}(\nu) B^{(mn)} =0,\;\;n=1,\dots,D.
\end{eqnarray}

\paragraph{Reduction 2.}
Let 
\begin{eqnarray}\label{red2}
&&
\tilde \Gamma^{(m)}(\lambda)=
\tilde \Gamma(\lambda) C^{(m)},\;\;C^{(1)}\equiv I_{n_0},\;\;C^{(n)} B^{(mn)}=B^{(m)},\;\;
m,n=1,\dots,D,
\end{eqnarray}
in addition to reduction (\ref{1st_order_ABP}).
Here $C^{(m)}$ and $B^{(m)}$  are some $n_0\times n_0$ constant  matrices and $\tilde \Gamma(\lambda)$ is $n_0\times n_0$ matrix function.
As a consequence, we obtain
\begin{eqnarray}\label{red22}
\tilde A^{(m)}(\lambda) =\tilde A(\lambda) C^{(m)},\;\;\tilde A(\lambda)=\alpha^{(1)}(\lambda,\mu)\star\tilde\Gamma(\mu),\;\;
V^{(m)}(t)=V(t)C^{(m)},\;\;\tilde \Gamma(\lambda)=\tilde \Gamma^{(1)}(\lambda).
\end{eqnarray}
Then the system of $D$ equations (\ref{Q_simple}) 
reduces to the following single PDE:
\begin{eqnarray}\label{Q_simple2}
&&
\sum_{m=1}^{D}\left(
V_{t_m} +V C^{(m)}V\right)B^{(m)}=0.
\end{eqnarray}
This equation is written in the Introduction, see eq.(\ref{nl1_intr0}), 
and may be referred to  as a multidimensional generalization of (1+1)-dimensional $C$-integrable $N$-wave equation  (\ref{nlin_cl_Nwave}).
Reduction (\ref{red2}) changes 
internal constraints (\ref{beta_m}) and (\ref{eta_m_Gamma}) as follows: 
\begin{eqnarray}\label{beta_m2}
&&
\left(\Gamma^{(m)}(\lambda,\nu)+\tilde \Gamma(\lambda) C^{(m)} P(\nu)\right)*\chi_0(\nu,\mu) + \Gamma^{(m)}(\lambda,\mu)=0,
\\\label{eta_m_Gamma2}
&&
T^{(m)}(\lambda,\mu)=T^{(1)}(\lambda,\nu)\star\left(\Gamma^{(m)}(\nu,\tilde\nu)+\tilde \Gamma C^{(m)}(\nu) P(\tilde\nu)\right)*\alpha^{(1)}(\tilde\nu,\tilde\mu)
\star(T^{(1)})^{-1}(\tilde\mu,\mu),\\\nonumber
&&
m=1,\dots,D.
\end{eqnarray}
In turn, external constraint  (\ref{constrain1_red1}) reduces to the following single equation:
\begin{eqnarray}\label{constrain1_red2}
\sum_{m=1}^{D}(A^{(m)}(\lambda,\nu) +\tilde A(\lambda) C^{(m)} P(\nu))*\tilde A(\nu) B^{(m)} =0.
\end{eqnarray}

\subsubsection{Second order nonlinear PDEs for the functions $V^{(m)}$, $m=1,\dots,D$}
{\bf Theorem 2.3.} In addition to eqs.(\ref{Psi},\ref{t}), we impose the following external constraint for the matrix functions  $A^{(m)}(\lambda,\mu)$ and $\tilde A^{(m)}(\lambda)$ (instead of constraint (\ref{constrain1})):
\begin{eqnarray}\label{constrain2}
&&
\sum_{m,n=1}^{D}A^{(m)}(\lambda,\nu)*A^{(n)}(\nu,\mu) *\tilde A^{(l)}(\mu) B^{(mnl)} =\\\nonumber
&&
\sum_{m,n}^{D}A^{(m)}(\lambda,\mu)*\tilde A^{(n)}(\mu) P^{(mnl)}+
\sum_{m=1}^{D}\tilde A^{(m)}(\lambda) P^{(ml)},
\end{eqnarray}
where $B^{(mnl)}$, $P^{(mnl)}$, and $P^{(ml)}$ are some constant $n_0\times n_0$  matrices.
Then $n_0\times n_0$ matrix functions $V^{(m)}(t)$
are solutions to the following system of nonlinear PDEs:
\begin{eqnarray}\label{2Q_V}
&&
\sum_{m,n=1}^{D} \left[
V^{(l)}_{t_nt_m} +(V^{(n)} V^{(l)})_{t_m} +V^{(m)}V^{(l)}_{t_n}+ V^{(m)}V^{(n)}V^{(l)} 
\right] B^{(mnl)}
+\\\nonumber
&&
\sum_{m,n=1}^{D} 
(V^{(n)}_{t_m}+V^{(m)}V^{(n)}) {\cal{A}}^{(mnl)}_1 
+
\sum_{m=1}^{D} V^{(m)} {\cal{A}}^{(ml)}_2=0,\;\;l=1,\dots,D,
\\\nonumber
&&
{\cal{A}}^{(mnl)}_1 = {\cal{A}}^{(l)}  B^{(mnl)} +P^{(mnl)},\;\;
{\cal{A}}^{(ml)}_2=
\sum_{n=1}^D {\cal{A}}^{(n)} P^{(mnl)}- P^{(ml)}.
\end{eqnarray}

{\bf Proof:} Applying operator  $*\tilde A^{(n)}$ to the eq.(\ref{lin})  from the right one gets 
eq.(\ref{2U}),
which introduces a new set of fields $U^{(mn)}$  defined by eq.(\ref{UW}). However, we may not eliminate these fields from the system (\ref{2U}) because constraint (\ref{constrain1}) is not valid in this theorem.
Instead of this, we 
 derive nonlinear PDEs for these fields as follows.
Applying operator $*A^{(n)}*\tilde A^{(l)}  $  to the eq.(\ref{lin})  from the right one gets
\begin{eqnarray}\label{2Q0}
&&
E^{(mnl)}(t):=
U^{(nl)}_{t_m}(t) +V^{(m)}(t)U^{(nl)}(t)+U^{(mnl)}(t) =0,
\end{eqnarray}
where one more set of matrix fields appears:
\begin{eqnarray}\label{U_mnl}
U^{(mnl)}(t)=(W(\lambda;t)-P(\lambda))*A^{(m)}(\lambda,\nu)* A^{(n)}(\nu,\mu)* \tilde A^{(l)}(\mu).
\end{eqnarray}
Due to the constraint (\ref{constrain2}), 
these fields may be eliminated in a proper 
combination of equations (\ref{2Q0}), namely,
$\sum_{m,n=1}^{D} E^{(mnl)} B^{(mnl)}$. Then, substituting $U^{(mn)}$ from eq.(\ref{2U}) one ends up with eq.(\ref{2Q_V}). $\blacksquare$

Emphasize that nonlinear equations (\ref{Q}) and (\ref{2Q_V}) do not represent commuting flows since we assume different external constraints (\ref{constrain1}) and (\ref{constrain2})  for matrix functions $A^{(m)}(\lambda,\mu)$ 
and $\tilde A^{(m)}(\lambda)$ deriving these equations. These two constraints  are not compatible in general.

\paragraph{Reduction 1.} Let
\begin{eqnarray}\label{red3}
&&
 P^{(mnl)}=-{\cal{A}}^{(l)} B^{(mnl)} ,\\\nonumber
&&
P^{(ml)}= \sum_{n=1}^D {\cal{A}}^{(n)} P^{(mnl)}.
\end{eqnarray}
Then eq.(\ref{2Q_V}) reads
\begin{eqnarray}\label{2Q_V_red}
&&
\sum_{m,n=1}^{D} \left[
V^{(l)}_{t_nt_m} +(V^{(n)} V^{(l)})_{t_m} +V^{(m)}V^{(l)}_{t_n}+ V^{(m)}V^{(n)}V^{(l)} 
\right] B^{(mnl)}=0,\;\;l=1,\dots,D.
\end{eqnarray}
Constraint (\ref{constrain2}) reduces to the following one:
\begin{eqnarray}\label{constrain3}
&&
\sum_{m,n=1}^{D}\Big(A^{(m)}(\lambda,\nu)*A^{(n)}(\nu,\mu) 
+A^{(m)}(\lambda,\nu)*\tilde A^{(n)}(\nu) P(\mu)+\\\nonumber
&&
\tilde A^{(m)}(\lambda)P(\nu)*\tilde A^{(n)}(\nu) P(\mu)\Big)*
\tilde A^{(l)}(\mu) B^{(mnl)} =0.
\end{eqnarray}

\paragraph{Reduction 2.}
Along with reduction (\ref{red3}) we consider reduction (\ref{red2},\ref{red22}) with
\begin{eqnarray}
C^{(l)} B^{(mnl)}=B^{(mn)},
\end{eqnarray}
where $B^{(mn)}$  are some constant matrices.
System (\ref{2Q_V_red}) reduces to the following  single PDE: 
\begin{eqnarray}\label{2Q_V_red2}
&&
\sum_{m,n=1}^{D} \left[
V_{t_n t_m} +(V C^{(n)} V )_{t_m} +V C^{(m)}V_{t_n}+
 V C^{(m)}V C^{(n)}V 
\right] B^{(mn)}=0.
\end{eqnarray}
This equation is written in the Introduction, see eq.(\ref{nl2_intr1}).
Internal constraints
(\ref{beta_m2}) and (\ref{eta_m_Gamma2}) remain valid for this case as well.
External constraint (\ref{constrain3}) reduces to the following single equation:
\begin{eqnarray}\label{constrain23}
&&
\sum_{m,n=1}^{D}\Big(A^{(m)}(\lambda,\nu)*A^{(n)}(\nu, \mu) 
+A^{(m)}(\lambda,\nu)*\tilde A (\nu)C^{(n)} P(\mu)+\\\nonumber
&&
\tilde A(\lambda) C^{(m)}P(\nu)*\tilde A(\nu) C^{(n)} P(\mu)\Big)
*\tilde A(\mu) B^{(mn)} =0.
\end{eqnarray}

\subsubsection{Solutions to the first order  nonlinear PDE (\ref{Q_simple2})}

The problem of richness of the available solution space will be considered for the nonlinear PDE (\ref{Q_simple2}). 
We show that solution space to this equation may be full provided that all constraints 
(\ref{beta_m2}),
(\ref{eta_m_Gamma2}) and (\ref{constrain1_red2}) may be resolved keepeng proper arbitrariness of functions $\tau^{(m)}(\nu)$. 
Examples of particular solutions will be considered as well.

\paragraph{On the dimensionality of the available solution space.}
We 
estimate the dimensionality of solution space for small $\chi$. In this case formula (\ref{Psi}) yields 
$W(\lambda;t)\approx P(\nu)*\chi(\nu,\lambda;t)$ and formula (\ref{Q_sol}) gives us 
\begin{eqnarray}\label{V_approx}
V(t)\approx \Big(P(\nu)*\chi(\nu,\lambda;t) - P(\lambda)\Big)*\tilde A(\lambda).
\end{eqnarray}
By construction, if all $\tau^{(m)}(\nu)$ ($m=1,\dots,D$) are arbitrary functions,   this expression preserves the following 
 arbitrary $n_0\times n_0$ matrix function of all $D$ variables:
\begin{eqnarray}\label{C_dim}
&&
F(t)=P*\alpha^{(1)}\star(T^{(1)})^{-1}\star \xi^{(1)}*\tilde A \equiv
\int g(\nu) e^{\sum_{i=1}^D \tau^{(i)}(\nu) t_i}g_2(\nu) d\Omega_2(\nu),\\\nonumber
&&
g_1(\nu)=P(\tilde\nu)*\alpha^{(1)}(\tilde\nu,\mu)\star(T^{(1)})^{-1}(\mu,\lambda)\star T(\lambda,\nu) ,\;\;
g_2(\nu)=T^{-1}(\nu,\tilde\nu)\star  \xi_0(\tilde\nu,\mu)*\tilde A (\mu)
\end{eqnarray}
 However, dimensionality of this function reduces due to the presence 
of constraints (\ref{beta_m2}),
(\ref{eta_m_Gamma2}) and (\ref{constrain1_red2}) which impose relations among $\tau^{(m)}(\nu)$. An important question is whether the dimensionality of the function (\ref{C_dim}) may be equal to $D-1$, which is necessary for fullness of the solution space. At first glance, we may expect the positive answer. In fact, eq.(\ref{beta_m2}) may be satisfied using special structures of $\Gamma^{(m)}$, $P$ and $\chi_0$, as it is done in the example below, see eqs.(\ref{parameters}). Next, eq.(\ref{eta_m_Gamma2}) relates $\tau^{(m)}$ with $\Gamma^{(m)}$ and $\tilde \Gamma^{(m)}$ which, in general, keeps arbitrariness of all $\tau^{(m)}(\nu)$ and consequently  does not restrict dimensionality of the above written arbitrary function. Finally, eq.(\ref{constrain1_red2}) 
must be considered as a single relation among $\tau^{(m)}(\nu)$, $m=1,\dots$, 
redusing the dimensionality of the function (\ref{C_dim}) from $D$ to $D-1$, which means the full dimensionality of the solution space. Thus, we may expect new completely integrable nonlinear PDEs in the derived class of equations. 

We have outlined a rough analysis of the solution space dimensionality. 
More detailed analysis must be carried out for particular equations and  remains beyond the scope of this paper.

\paragraph{Construction of explicite solutions.}
\label{Section:first_order}
Now we derive a family of particular solutions in the form of rational functions of exponents. Solitons and kinks are the most famous representatives of this family. To derive such solutions,
we take
\begin{eqnarray}\label{Omega}
&&
d\Omega_1(\lambda) = \sum_{i=1}^M\delta(\lambda-a_i)d\lambda,\;\; 
d\Omega_2(\lambda) = \sum_{i=1}^N\delta(\lambda-b_i)d\lambda,\;\;T(\lambda,\mu)={\cal{I}}_2(\lambda,\mu),\\\nonumber
&&
{\cal{I}}_1\to I_{Mn_0},\;\;{\cal{I}}_2\to I_{Nn_0}.
\end{eqnarray}
Then all integral equations reduce to the algebraic ones. 
We use notations
\begin{eqnarray}\label{notations_discr}
&&
\hat W=\left[
\begin{array}{ccc}
W(a_1)&\cdots & W(a_M)
\end{array}
\right],\;\;\hat \chi=\left[
\begin{array}{ccc}
\chi(a_1,a_1)&\cdots & \chi(a_1,a_M)\cr
\cdots &\cdots &\cdots \cr
\chi(a_M,a_1)&\cdots & \chi(a_M,a_M)
\end{array}
\right],\\\nonumber
&&
\hat \chi_0=\left[
\begin{array}{ccc}
\chi_0(a_1,a_1)&\cdots & \chi_0(a_1,a_M)\cr
\cdots &\cdots &\cdots \cr
\chi_0(a_M,a_1)&\cdots & \chi_0(a_M,a_M)
\end{array}
\right],\;\;
\hat \alpha^{(1)}=\left[
\begin{array}{ccc}
\alpha^{(1)}(a_1,b_1)&\cdots & \alpha^{(1)}(a_1,b_N)\cr
\cdots &\cdots &\cdots \cr
\alpha^{(1)}(a_M,b_1)&\cdots & \alpha^{(1)}(a_M,b_N)
\end{array}
\right],\\\nonumber
&&
\hat \Gamma^{(m)}=\left[
\begin{array}{ccc}
\Gamma^{(m)}(b_1,a_1)&\cdots & \Gamma^{(m)}(b_1,a_M)\cr
\cdots &\cdots &\cdots \cr
\Gamma^{(m)}(b_N,a_1)&\cdots & \Gamma^{(m)}(b_N,a_M)
\end{array}
\right],\;\;\hat{\tilde\Gamma}=\left[
\begin{array}{ccc}
\tilde\Gamma(b_1)\cr
\cdots \cr
\tilde \Gamma(b_N)
\end{array}
\right],\\\nonumber
&&
\hat \xi_0=\left[
\begin{array}{ccc}
\xi_0(b_1,a_1)&\cdots & \xi_0(b_1,a_M)\cr
\cdots &\cdots &\cdots \cr
\xi_0(b_N,a_1)&\cdots & \xi_0(b_N,a_M)
\end{array}
\right],\;\;\;\;\hat P=[P(a_1)\;\;\cdots\;\;P(a_M)],\\\nonumber
&&
\hat \tau^{m)}={\mbox{diag}}(\tau^{(m)}(b_1),\cdots,\tau^{(m)}(b_N)).
\end{eqnarray} 
Solution $V$ is given by eq.(\ref{Q_sol}) together with reduction (\ref{red2},\ref{red22}) as follows:
\begin{eqnarray}\label{V_sol0}
V=(\hat W -\hat P)\hat{\tilde A}
\end{eqnarray}
 where $\hat W$ is solution to eq.(\ref{Psi}):
\begin{eqnarray}\label{W_sol}
\hat W=\hat P
\hat \chi(\hat \chi+I_{Mn_0})^{-1}
\end{eqnarray}
Substituting eq.(\ref{W_sol}) into eq.(\ref{V_sol0}) we obtain 
\begin{eqnarray}\label{V_sol}
&&
V=
\hat P\Big(
\hat \chi(\hat \chi+I_{Mn_0})^{-1}-I_{Mn_0}\Big) \hat \alpha^{(1)}\hat{\tilde \Gamma}.
\end{eqnarray}
Since
$\chi$ is given by eq.(\ref{chi}), one has
\begin{eqnarray}\label{hat_chi}
&&
\hat \chi=\hat \alpha^{(1)}(\hat \tau^{(1)})^{-1} e^{\sum_{i=1}^{D} \hat \tau^{(i)} t_i} \hat \xi_0+\hat \chi_0,
\end{eqnarray}
where we substitute eq.(\ref{xi_expl}) for $\xi^{(1)}$.

Matrices  $\hat \Gamma^{(m)}$, $\hat{\tilde\Gamma}$ and $\hat\tau^{(m)}$ must satisfy  constraints (\ref{beta_m2}),
(\ref{eta_m_Gamma2}) and (\ref{constrain1_red2}),
which read in our case, $m=1,\dots,D$:
\begin{eqnarray}\label{beta_m2_sol}
&&
\left(\hat \Gamma^{(m)}+\hat{\tilde \Gamma} C^{(m)} \hat P\right)\chi_0 + \hat \Gamma^{(m)}=0,
\\\label{eta_m_Gamma2_sol}
&&
\hat \tau^{(m)}=\hat \tau^{(1)}\left(\hat \Gamma^{(m)}+\hat{\tilde \Gamma} C^{(m)} \hat P\right)\hat \alpha^{(1)}
(\hat \tau^{(1)})^{-1},\\
\label{constrain1_red2_sol}
&&
\sum_{m=1}^{D}\hat \alpha^{(1)}(\hat \Gamma^{(m)} +\hat{\tilde\Gamma} C^{(m)} \hat P)\hat \alpha^{(1)}
\hat{\tilde \Gamma} B^{(m)} =0.
\end{eqnarray}

Analisis of eqs.(\ref{beta_m2_sol}) points on  two different types of solutions to them. First type is assotiated with $\det(\hat \chi_0+I_{M n_0}) \neq 0$. Then eqs.(\ref{beta_m2_sol})
may be solved for $\Gamma^{(m)}$ and 
one can show that  multidimensional PDE (\ref{Q_simple2}) may be splitted into 
 a set of independent compatible   Ordinary Differential Equations (ODEs). We will not consider this case. 
Second type  is assotiated with $\det(\hat \chi_0+I_{M n_0})= 0$ and leads to truly multidimensional solutions to  eq.(\ref{Q_simple2}). Namely this case is considered hereafter.

Looking for the particular solutions to eq.(\ref{beta_m2_sol})  we decompose it into two equations:
\begin{eqnarray}\label{bet_m2_sol}
\hat P\hat \chi_0=0,\;\;\hat \Gamma^{(m)} (\hat \chi_0+I_{M n_0}) =0,
\end{eqnarray}
which means that the rows of $\hat P$ and $\hat \Gamma^{(m)}$ are orthogonal to the columns of 
$\hat \chi_0$ and   $\hat \chi_0+I_{Mn_0}$ respectively.

Note that eq.(\ref{eta_m_Gamma2_sol}) with $m=1$ reads:
\begin{eqnarray}
\label{eta_m_Gamma2_sol_tr}
&&
\hat \tau^{(1)}=\left(\hat \Gamma^{(1)}+\hat{\tilde \Gamma}  \hat P\right)\hat \alpha^{(1)}.
\end{eqnarray}
Since, $\hat \tau^{(1)}$ must be invertable, we 
require $M>N$. Then eq.(\ref{constrain1_red2_sol}) may be simplified   removing
 $\hat \alpha^{(1)}$ as a left factor in this equation:
\begin{eqnarray}\label{constrain1_2}
&&
\sum_{m=1}^{D}  (\hat \Gamma^{(m)}+\hat{\tilde \Gamma} C^{(m)} \hat P)\hat\alpha^{(1)} \hat{\tilde \Gamma} B^{(m)} =
0.
\end{eqnarray}

\paragraph{Simple example of solution.}
We consider the three-dimensional nonlinear PDE (\ref{Q_simple2}), i.e. $D=3$:
\begin{eqnarray}\label{Q_simple2_3}
&&
\sum_{m=1}^{3}\left(
V_{t_m} +V C^{(m)}V\right)B^{(m)}=0.
\end{eqnarray}
Let  $n_0=2$, $M=5$, $N=2$,
\begin{eqnarray}
B^{(1)}=C^{(1)}=I_{2},\;\;B^{(i)}={\mbox{diag}}(b^{(i)}_1,b^{(i)}_2),\;\;
C^{(i)}={\mbox{diag}}(c^{(i)}_1,c^{(i)}_2),\;\;i=2,3.
\end{eqnarray}
In order to satisfy eqs.(\ref{eta_m_Gamma2_sol}), (\ref{bet_m2_sol})
and (\ref{constrain1_2}) we take the following matrices 
$\hat \chi_0$, $\hat P$, $\hat \Gamma^{(m)}$, $\hat{\tilde\Gamma}$:
\begin{eqnarray}\label{parameters}
&&
\hat \chi_0=\left[
\begin{array}{cc}
-I_{2} & Z_{2,8}\cr
F_{2, 2} &F_{2,8 }\cr
Z_{6,2} & Z_{6,8}
\end{array}
\right],\;\;
F_{2,2}=Z,\;\;F_{2,8}=[J_4 \;Z \;J_5 \;Z],
\\\nonumber
&&
\hat \alpha^{(1)}=\left[
\begin{array}{c}
I_4\cr I_4\cr
J_0
\end{array}
\right],\;\;\hat \xi_0=(\alpha^{(1)})^T
\\\nonumber
&&
\hat P=\left[Z \;Z\;Z\;J_1\;Z\right],\;\;\\\nonumber
&&
\hat \Gamma^{(1)}=\left[
\begin{array}{ccccc}
J_2 &Z&Z&Z&Z\cr
Z&Z&Z&Z&Z
\end{array}
\right],\;\;\hat \Gamma^{(2)}=\hat \Gamma^{(3)}=Z_{4,10},\\\nonumber
&&
\hat{\tilde\Gamma}\equiv\hat{\tilde \Gamma}^{(1)}=\left[
\begin{array}{cc}
Z\cr J_3
\end{array}
\right],\;\;\hat \tau^{(1)}={\mbox{diag}}(1,2,3,4),\;\;
\hat \tau^{(2)}={\mbox{diag}}(0,0,3 c^{(2)}_2,4c^{(2)}_1),\\\nonumber
&&\hat \tau^{(3)}={\mbox{diag}}(0,0,3 c^{(3)}_2,4c^{(3)}_1).
\end{eqnarray}
In addition, we obtain expressions for $c^{(3)}_i$, $i=1,2$:
\begin{eqnarray}\label{c3}
 c^{(3)}_i=-\frac{1+b^{(2)}_i c^{(2)}_i}{b^{(3)}_i},\;\;i=1,2.
\end{eqnarray}
Here $Z_{i,j}$ and $Z$ are $i\times j$  and $2\times 2$ zero matrices respectively,
\begin{eqnarray}
J_0=[I_2\;\;Z],\;\;
J_1=\left[
\begin{array}{cc}
0&1\cr
1&0
\end{array}
\right],\;\;J_2=\left[
\begin{array}{cc}
1&0\cr
0&2
\end{array}
\right],\;\;J_3=\left[
\begin{array}{cc}
0&3\cr
4&0
\end{array}
\right],\;\;
\end{eqnarray}
\begin{eqnarray}
&&
J_4=\left[
\begin{array}{cc}\displaystyle
\frac{8 p_3 + p_2 (6 - p_3) - 4 p_4 - p_1 (12 - p_4)}{
   2 (6 p_1 - 3 p_2 - 2 p_3 + p_4)}&1\cr\displaystyle
\frac{(-12 + 3 p_2 - p_4) (-2 p_3 + p_4)}{
   12 (6 p_1 - 3 p_2 - 2 p_3 + p_4)}&0
\end{array}
\right],\\\nonumber
&&
J_5=\left[
\begin{array}{cc}
0&\displaystyle
1+\frac{4(3 p_1 -  p_3 - 6)}{
    12 - 3 p_2 + p_4}
\cr\displaystyle
-1+\frac{p_2}{4} -\frac{p_4}{12}&0
\end{array}
\right],\;\;
\end{eqnarray}
Diagonal elements of  matrices $B^{(i)}$, $i=2,3$, and $C^{(2)}$ remain arbitrary.
Substituting eqs.(\ref{parameters}) into eq.(\ref{V_sol},\ref{hat_chi}) one obtains $V$ as a  rational expression of exponents:
\begin{eqnarray}
&&
V(t)=\frac{1}{D}\left[
\begin{array}{cc}\displaystyle
 {-4 (e^{ \eta_2}  p_2 +  p_4)}
&\displaystyle
 {f_{12}e^{ \eta_1}}\cr\displaystyle
{ f_{21}e^{ \eta_2}}&\displaystyle
{ -3 (e^{\eta_1}  p_3 +   p_4)}
\end{array}
\right],\\\nonumber
&&
D={
 e^{ \eta_1+\eta_2}  p_1 +  e^{\eta_2}  p_2 + 
     e^{\eta_1}  p_3 + p_4}
,\\\nonumber
&&
f_{12}=-3\frac{(-12 + 3  p_2 -  p_4) 
    ( p_2  p_3 -  p_1  p_4)}{4 (-12  p_1 + 6  p_2 + 4  p_3 - 
     2  p_4)},\;\;
f_{21}=16\frac{12  p_1 - 6  p_2 - 4  p_3 + 2  p_4}{-12 + 3  p_2 -  p_4}.
\end{eqnarray}
Here
\begin{eqnarray}
&&
\eta_1=4 t_1 + 4 c^{(2)}_1 t_2- \frac{4}{b^{(3)}_1}(1+b^{(2)}_1 c^{(2)}_1)t_3,\\\nonumber
&&
\eta_1=3 t_1 + 3 c^{(2)}_2 t_2-\frac{3}{b^{(3)}_2}(1+b^{(2)}_2 c^{(2)}_2)t_3.
\end{eqnarray}
We see that all  elements of $V$ are kinks. 

Relations (\ref{c3}) show that not all  matrix coefficients in eq.(\ref{Q_simple2_3}) are arbitrary. They are related by the following equation:
\begin{eqnarray}\label{BC}
C^{(3)} B^{(3)}+C^{(2)} B^{(2)}+I_2=0.
\end{eqnarray}
Thus we have constructed particular solution to the three-dimensional  nonlinear PDE (\ref{Q_simple2_3}) with diagonal matrices $C^{(i)}$, $B^{(i)}$ related by eq.(\ref{BC}).

\section{Multidimensional generalization of   $S$-integrable PDEs}
\label{Section:Sint}

\subsection{Starting equations}
\label{S_Section:general}
Algorithm developed in this section is based on 
 the same equation (\ref{Psi}) with different  function $\chi(t)$, which is defined  by the following system of equations 
\begin{eqnarray}\label{S_t}
\chi_{t_m}(\lambda,\mu;t)&=&\Big(
A^{(m)}(\lambda,\nu)+\tilde A^{(m)}(\lambda) P(\nu)\Big)* \chi(\nu,\mu;t) -\chi(\lambda,\nu;t) *A^{(m)}(\nu,\mu),\\\nonumber
&&
m=1,\dots,D
\end{eqnarray} 
instead of system (\ref{t}).
Here, again, $A^{(m)}(\lambda,\nu)$ and $\tilde A^{(m)}(\lambda)$ are $n_0\times n_0$   matrix functions of arguments.

Matrices $A^{(m)}$ and $\tilde A^{(m)}$  have to provide compatibility of 
system (\ref{S_t}). Similar to  eq.(\ref{t}), there are two different methods that provide this compatibility. The first one yields the classical $S$-integrable nonlinear PDEs,
 Sec.\ref{S_Section:classical}, 
while the second method  yields a new type of nonlinear PDEs whose complete integrability is not clarified yet, Sec.\ref{S_Section:gen}. However, our algorithm supplyes, at least, a rich manifold of particular solutions to these PDEs.

\subsection{First method: classical $S$-integrable (2+1)-dimensional $N$-wave equation}
\label{S_Section:classical}
Consider the compatibility condition of eqs.(\ref{S_t}) in the following form:
\begin{eqnarray}\label{S_compatA}
&&
\left(A^{(m)}(\lambda,\nu) +\tilde A^{(m)}(\lambda) P(\nu) \right)*\chi_{t_n}(\nu,\mu;t) -\chi_{t_n}(\lambda,\nu;t) *A^{(m)}(\nu,\mu)  =\\\nonumber
&&
\left(A^{(n)}(\lambda,\nu) +\tilde A^{(n)}(\lambda) P(\nu) \right)*\chi_{t_m}(\nu,\mu;t)-\chi_{t_m}(\lambda,\nu;t)*A^{(n)}(\nu,\mu),\;\;\forall \; n,m 
\end{eqnarray}
Substituting eq.(\ref{S_t}) for derivatives of $\chi$ we reduce eq.(\ref{S_compatA})
to the following one:
\begin{eqnarray}\label{S_LL}
&&
(L^{(m)}*L^{(n)}-L^{(n)}*L^{(m)})*\chi - \chi* (A^{(m)}* A^{(n)}-A^{(n)}* A^{(m)}) =0,\\\nonumber
&&
L^{(m)}(\lambda,\mu)=A^{(m)}(\lambda,\mu) +\tilde A^{(m)}(\lambda) P(\mu).
\end{eqnarray}
Since eq.(\ref{S_LL}) must  be valid for any function $\chi(t)$ (which is a solution to the system (\ref{S_t})), it is equivalent to two following equations relating matrix functions $A^{(m)}$, $\tilde A^{(m)}$ and $P$:
\begin{eqnarray}\label{S_LA1}
&&
A^{(m)}(\lambda,\nu) *A^{(n)}(\nu,\mu) -A^{(n)}(\lambda,\nu) *A^{(m)}(\nu,\mu)=0,\\\nonumber
&&
L^{(m)}*L^{(n)}-L^{(n)}*L^{(m)}=0\;\;\;\stackrel{{\mbox{eq.\ref{S_LA1}}}}{\Rightarrow} \\\label{S_LA3}\label{S_LA2}
&&
L^{(m)}(\lambda,\nu)* \tilde A^{(n)}(\nu)  P(\mu)  -
L^{(n)}(\lambda,\nu)* \tilde A^{(m)}(\nu)  P(\mu)  = \\\nonumber
&&
\tilde A^{(n)}(\lambda) P(\nu)* A^{(m)}(\nu,\mu)  - \tilde A^{(m)}(\lambda) P(\nu)*A^{(n)}(\nu,\mu).
\end{eqnarray}
In order to satisfy  eq.(\ref{S_LA3}) we require the following representation of $\tilde A^{(m)}(\lambda)$:
\begin{eqnarray}
\tilde A^{(m)}(\lambda) = \tilde A(\lambda) B^{(m)}, \;\;[B^{(m)},B^{(n)}]=0,
\end{eqnarray}
where $\tilde A(\lambda)$ and  $B^{(m)}$ are $n_0\times n_0$ matrix function and  constant matrix respectively.
Then eq.(\ref{S_LA3}) is equivalent to the following system:
\begin{eqnarray}\label{S_LA_AL}
L^{(m)}(\lambda,\nu) *\tilde A(\nu) B^{(n)} -L^{(n)}(\lambda,\nu)* \tilde A(\nu) B^{(m)} = 0,\\\label{S_LA_AL2}
B^{(n)} P(\nu) *A^{(m)}(\nu,\mu) -
 B^{(m)} P(\nu) *A^{(n)}(\nu,\mu) =0.
\end{eqnarray}
Eqs.(\ref{S_LA1},\ref{S_LA_AL},\ref{S_LA_AL2}) represent three constraints for  matrices $A^{(m)}$ and $\tilde A$.

{\bf Theorem 3.1.} 
 Let the matrix function 
$W(\lambda;t)$ be obtained as a solution to the integral equation (\ref{Psi}) with $\chi$ defined by 
eqs.(\ref{S_t}) supplemented with constraints (\ref{S_LA1},\ref{S_LA_AL},\ref{S_LA_AL2}). Then\newline

1. Function $W(\lambda;t)$ satisfies the following  system of compatible linear equations
\begin{eqnarray}\label{S_lin_cl}
E^{(nm)}(\mu;t)&:=&B^{(n)}\left( W_{t_m}(\mu;t) + V(t) B^{(m)} W(\mu;t) 
+W(\nu;t) *A^{(m)}(\nu,\mu) \right)-\\\nonumber
&&
B^{(m)}\left( W_{t_n}(\mu;t) +V(t)B^{(n)} W (\mu;t)
+W (\nu;t)*A^{(n)}(\nu,\mu) \right)=0 ,\\\label{S_Q_sol}\label{S_V}
&&
V(t)=(W(\nu;t) -P(\nu))*\tilde A(\nu)
,\\\nonumber
&&
n,m=1,\dots,D.
\end{eqnarray}
2. Matrix field $V(t)$, given by eq.(\ref{S_V}), satisfies the following $S$-integrable $N$-wave equation:
\begin{eqnarray}
\label{S_Nwave}
&&
\sum_{perm(n,m,l)}\Big( B^{(n)}( V_{t_m}  + V  B^{(m)}  V) -
 B^{(m)}( V_{t_n}  + V  B^{(n)}  V)\Big)B^{(l)}=0,
\end{eqnarray}
where $perm(n,m,l)$ means clockwise circle permutations.

{\bf Proof:} 
1. To derive eq.(\ref{S_lin_cl}), we 
differentiate eq.(\ref{Psi}) with respect to $t_m$.  Then, in view of eq.(\ref{S_t}), 
one gets the following equation:
\begin{eqnarray}\label{S_lin_1}
&& 
{\cal{E}}^{(m)}(\mu;t):=P(\lambda)*A^{(m)}(\lambda,\nu)*\chi(\nu,\mu;t)=
E^{(m)}(\nu;t)*\Big(\chi(\nu,\mu)+{\cal{I}}_1(\nu,\mu)\Big) ,\\\nonumber
&&
E^{(m)}(\mu;t)=W_{t_m}(\mu;t) + V^{(m)}(t) W(\mu;t) 
+W(\nu;t) *A^{(m)}(\nu,\mu).
\end{eqnarray}
Due to the constraint (\ref{S_LA_AL2}),  LHS in eqs.(\ref{S_lin_1}) may be removed using  the following combination of these equations:
\begin{eqnarray}
&&
B^{(n)}{\cal{E}}^{(m)}-B^{(m)}{\cal{E}}^{(n)} \;\;\Rightarrow\\\label{BcalE}
&&
(B^{(n)}E^{(m)}(\nu;t)-B^{(m)}E^{(n)}(\nu;t))*(\chi(\nu,\mu;t)+{\cal{I}}_1(\nu,\mu))=0.
\end{eqnarray}
Since operator $*(\chi(\nu,\mu)+{\cal{I}}_1(\nu,\mu))$ is invertable, one gets
\begin{eqnarray}
 {\cal{E}}^{(nm)}:=B^{(n)} E^{(m)}-B^{(m)} E^{(n)}=0,
\end{eqnarray} 
which coinsides with eq.(\ref{S_lin_cl}).

2. In order to derive  nonlinear PDE (\ref{S_Nwave})  we 
consider the following combination of eqs.(\ref{S_lin_cl}): 
\begin{eqnarray}\label{S_nlin_cl}
&&
 \sum_{perm(n,m,l)} E^{(nm)} *\tilde A  B^{(l)},
\end{eqnarray}
which yields eq.(\ref{S_Nwave}) in view of constraint 
(\ref{S_LA_AL}).
$\blacksquare$

$S$-integrable PDEs will not be considered in this paper.

\subsection{Second method: new class of nonlinear PDEs}
\label{S_Section:gen}

 We will use double indexes hereafter in this section, i.e.
\begin{eqnarray}
V^{(m)}\equiv V^{(m_1m_2)},\;\;t_m\equiv t_{m_1m_2},\;\;\sum_{m=1}^D f^{(m)}
\equiv \sum_{m_2=1}^{D_2} \sum_{m_1=1}^{D_1} f^{(m_1m_2)},\;\;\forall\;f,\;\;D\equiv (D_1,D_2),
\end{eqnarray}
and notation  $f^{(1)}\equiv f^{(11)}$, $\forall f$, for the sake of brevity. For instance, $\tau^{(1)}\equiv \tau^{(11)}$,  $\eta^{(1)}\equiv \eta^{(11)}$ and so on.
Similar to Sec.\ref{Section:Cint_new}, we will use representations (\ref{S_A_alp_bet}) for the matrix functions
 $A^{(m)}(\lambda,\mu)$ and $\tilde A^{(m)}(\lambda)$ and write eq.(\ref{S_t}) in the following form:
\begin{eqnarray}
\label{S_t_xi}
\chi_{t_m}(\lambda,\mu;t)&=&\alpha^{(m)}(\lambda,\nu)\star \xi^{(m)}(\nu,\mu)-
\bar\xi^{(m)}(\lambda,\nu) \star\beta^{(m)}(\nu,\mu),\\\label{S_A_xi}
&&
\xi^{(m)}(\lambda,\mu)=\left(\beta^{(m)}(\lambda,\nu) +\tilde \beta^{(m)}(\lambda) P(\nu)\right)*\chi(\nu,\mu;t),\\
\label{S_A_bxi}
&&
\bar\xi^{(m)}(\lambda,\mu) = \chi(\lambda,\nu) *\alpha^{(m)}(\nu,\mu),\;\;m_i=1,\dots,D_i,\;\;i=1,2.
\end{eqnarray}
Then the compatibility condition for system (\ref{S_t}) is equivalent to the 
compatibility condition for  system (\ref{S_t_xi}) which reads
\begin{eqnarray}\label{comp_xi}
\alpha^{(m)}\star \xi^{(m)}_{t_n}-\bar\xi^{(m)}_{t_n}\star \beta^{(m)} = \alpha^{(n)}\star\xi^{(n)}_{t_m}-
\bar\xi^{(n)}_{t_m}\star \beta^{(n)},
\;\;\;\forall \;n,m.
\end{eqnarray}
To satisfy this condition we, first, assume that
 $\xi^{(m)}$ and $\bar \xi^{(m)}$ are expressed in terms of 
$\xi^{(1)}$ and $\bar \xi^{(1)}$ as follows:
\begin{eqnarray}
\label{S_alpha_xi1}
&&\xi^{(m)}(\lambda,\mu;t)=\eta^{(m)}(\lambda,\nu)\star\xi^{(1)}(\nu,\mu),\;\;m_1+m_2>2,\;\;
\eta^{(1)}(\lambda,\nu)={\cal{I}}_2(\lambda,\nu),\\
\label{S_beta_xi1}
&&
\bar\xi^{(m)}(\lambda,\mu;t)=\bar\xi^{(1)}(\lambda,\nu;t)\star\bar\eta^{(m)}(\nu,\mu),\;\;
m_1+m_2>2,\;\;\bar\eta^{(1)}(\nu,\mu)={\cal{I}}_2(\lambda,\nu),
\end{eqnarray}
where 
 $\eta^{(m)}(\nu,\mu)$ and  $\bar\eta^{(m)}(\nu,\mu)$  are some  $n_0\times n_0$ matrix functions which will be specified below.
Second, we define $\xi^{(1)}_{t_m}$ and $\bar \xi^{(1)}_{t_m}$ in terms of $\xi^{(1)}$ and 
$\bar\xi^{(1)}$ as follows:
\begin{eqnarray}
\label{S_alpha_xi2}
\xi^{(1)}_{t_m}(\lambda,\mu;t) &=& T^{(m)}(\lambda,\nu)\star \xi^{(1)}(\nu,\mu;t) 
,\\
\label{S_beta_xi2}
\bar\xi^{(1)}_{t_m}(\lambda,\mu;t) &=&\bar\xi^{(1)}(\lambda,\nu;t)\star \bar T^{(m)}(\nu,\mu),
\end{eqnarray}
where $T^{(m)}(\lambda,\nu)$ and $\bar T^{(m)}(\lambda,\nu)$ are $n_0\times n_0$ matrix functions.
At last, substitute eqs.(\ref{S_alpha_xi1}-\ref{S_beta_xi2}) into eq.(\ref{comp_xi}). Since resulting equation must be valid for any possible $\xi^{(1)}$ and $\bar \xi^{(1)}$, we get the following expressions for $\alpha^{(m)}$ and $\beta^{(m)}$:
\begin{eqnarray}
\label{S_alpha_m}
&&
\alpha^{(m)}(\lambda,\mu)=\Big(\alpha^{(1)}(\lambda,\nu) * T^{(m)}(\nu,\tilde\nu) \star (T^{(1)})^{-1}(\tilde\nu,\nu) \Big)\star(\eta^{(m)})^{-1}(\nu,\mu),
\\\label{S_beta_m}
&&
\beta^{(m)}(\lambda,\mu)= (\bar\eta^{(m)})^{-1}(\lambda,\nu)\star (\bar T^{(1)})^{-1}(\nu,\tilde\nu)\star
 \bar T^{(m)}(\tilde \nu,\tilde\mu)\Big)\star
\beta^{(1)}(\tilde\mu,\mu).
\end{eqnarray}
In turn, compatibility  of eqs.(\ref{S_alpha_xi2}) and (\ref{S_beta_xi2}) requires
\begin{eqnarray}\label{com_T_bT}
&&
T^{(m)}\star T^{(n)}-T^{(n)}\star T^{(m)}
=0\;\;\Rightarrow \;\;T^{(m)}(\lambda,\mu)=\left(T(\lambda,\nu) \tau^{(m)}(\nu)\right)\star T^{-1}(\nu,\mu),
\\\nonumber
&&
\bar T^{(m)}\star\bar  T^{(n)}-\bar T^{(n)}\star \bar T^{(m)}
=0\;\;\Rightarrow \;\;\bar T^{(m)}(\lambda,\mu)=\bar T^{-1}(\lambda,\nu) \star
\left(\bar\tau^{(m)}(\nu) T^{-1}(\nu,\mu)\right),\\\nonumber
&&
[\tau^{(m)}(\nu),\tau^{(n)}(\nu)]=0,\;\;[\bar\tau^{(m)}(\nu),\bar\tau^{(n)}(\nu)]=0,
\end{eqnarray} 
where $T(\lambda,\nu)$, $\bar T(\lambda,\nu)$,  $ \tau^{(m)}(\nu)$ and $ \bar\tau^{(m)}(\nu)$ are $n_0\times n_0$ matrix functions
(compare with eq.(\ref{com_1})). Thus, compatibility condition of eqs.(\ref{S_t_xi}) generates eqs.(\ref{S_alpha_xi1}-\ref{com_T_bT}).

Now we may integrate eqs.(\ref{S_alpha_xi2}) and (\ref{S_beta_xi2})  obtaining the following expressions for $\xi^{(1)}$ and $\bar\xi^{(1)}$:
\begin{eqnarray}
\\\label{S_xi_t}
&&
\xi^{(1)}(\lambda,\mu;t)=\left(T(\lambda,\nu)e^{\sum_{i=1}^{D} \tau^{(i)}(\nu) t_i}\right)\star T^{-1}(\nu,\tilde\nu)\star \xi_0(\tilde\nu,\mu),
,\\
\label{S_bar_xi_t}
&&
\bar\xi^{(1)}(\lambda,\mu;t)=\bar\xi_0(\lambda,\nu)\star \bar T^{-1}(\nu,\tilde\nu)\star \left(
e^{\sum_{i=1}^{D} \bar\tau^{(i)}(\tilde\nu) t_i} \bar T(\tilde\nu,\mu)\right).
\end{eqnarray}
Here  $\xi_0(\tilde\nu,\mu)$ and  $\bar \xi_0(\lambda,\nu)$   are $n_0\times n_0$   matrix functions.
Finally, integrating  eq.(\ref{S_t_xi}) with $m_1=m_2=1$,  one obtains 
\begin{eqnarray}\label{S_chi}
\chi(\lambda,\mu;t)=\alpha^{(1)}(\lambda,\nu)\star (T^{(1)})^{-1}(\nu,\tilde\nu)\star \xi^{(1)}(\tilde\nu,\mu)- \bar\xi^{(1)}(\lambda,\nu)\star (\bar T^{(1)})^{-1}(\nu,\tilde\nu) \star\beta^{(1)}(\tilde\nu,\mu).
\end{eqnarray}

It is convenient  to rewrite eqs.(\ref{S_A_alp_bet}) using  eqs.(\ref{S_alpha_m}) and (\ref{S_beta_m}) as follows:
\begin{eqnarray}
\label{S_A_alp_bet_mod}
A^{(m)}(\lambda,\mu)&=&\alpha^{(1)}(\lambda,\nu)\star\gamma^{(m)}(\nu,\tilde \nu)\star\beta^{(1)}(\tilde\nu,\mu),\\
\label{S_A_alp_bet_mod2}
\tilde A^{(m)}(\lambda)&=&\alpha^{(1)}(\lambda,\nu)\star\tilde\Gamma^{(m)}(\nu),\;\;\forall\;m,
\end{eqnarray}
where
\begin{eqnarray}
\label{S_A_alp_bet_mod_gamma}
\gamma^{(m)}(\lambda,\mu)&=&T^{(m)}(\lambda,\nu)\star (T^{(1)})^{-1}(\nu,\tilde\nu)  (\eta^{(m)})^{-1}(\tilde\nu,\tilde\mu)\star
\\
\nonumber
&&
(\bar\eta^{(m)})^{-1}(\tilde\mu,\bar\mu)\star (\bar T^{(1)})^{-1}(\bar\mu,\tilde\lambda) \star \bar T^{(m)}(\tilde\lambda,\mu),
\\
\label{S_A_alp_bet_mod2_tGamma}
\tilde\Gamma^{(m)}(\lambda)&=&T^{(m)}(\lambda,\nu) \star (T^{(1)})^{-1}(\nu,\tilde\nu)\star
  (\eta^{(m)})^{-1}(\tilde\nu,\tilde\mu)*
\tilde\beta^{(m)}(\tilde\mu).
\end{eqnarray} 
Representations (\ref{S_A_alp_bet_mod},\ref{S_A_alp_bet_mod2}) will be used in the rest of Sec.\ref{S_Section:gen}.

\subsubsection{Internal constraints for $\alpha^{(1)}$, $\beta^{(1)}$, $\tilde\Gamma^{(m)}$,
$T$, $\bar T$, $\tau^{(m)}$ and  $\bar \tau^{(m)}$}
\label{Section:S_internal}
Note that the definitions of $\xi^{(m)}$ and $\bar\xi^{(m)}$ in terms of $\chi$, i.e. eqs.(\ref{S_A_xi}) and (\ref{S_A_bxi}),
must be consistent with eq.(\ref{S_chi}). This requirement  generates some constraints for $\alpha^{(1)}$, $\beta^{(1)}$, $\tilde\Gamma^{(m)}$,
$T$, $\bar T$, $\tau^{(m)}$ and  $\bar \tau^{(m)}$. To derive constraints assotiated with definition (\ref{S_A_xi}), we 
apply operator $(\beta^{(m)}+\tilde\beta^{(m)}P)*$  to the eq.(\ref{S_chi})  from the left obtaining the following equation:
\begin{eqnarray}\label{xi_interm}
\xi^{(m)}(\lambda,\mu;t)&=&(\beta^{(m)}(\lambda,\nu)+\tilde\beta^{(m)}(\lambda)P(\nu))*\alpha^{(1)}(\nu,\tilde\nu)\star (T^{(1)})^{-1}(\tilde\nu,\tilde\mu)\star \xi^{(1)}(\tilde\mu,\mu) -\\\nonumber
&& (\beta^{(m)}(\lambda,\nu)+\tilde\beta^{(m)}(\lambda)P(\nu))\star \bar\xi^{(1)}(\nu,\tilde\nu)*
(T^{(1)})^{-1}(\tilde\nu,\tilde\mu)\star \beta^{(1)}(\tilde\mu,\mu).
\end{eqnarray}
Substitute eq.(\ref{S_alpha_xi1}) for $\xi^{(m)}$ and require that resulting equation is identity for any $\xi^{(1)}$.  Then eq.(\ref{xi_interm}) becomes decomposed into two following constraints:
\begin{eqnarray}\label{bxi_00}\label{bar_xi_00}
&&
(\beta^{(m)}(\lambda,\nu)+\tilde\beta^{(m)}(\lambda) P(\nu) )*\bar\xi_0(\nu,\mu)=0,
\\\label{S_eta_m}
&&
\eta^{(m)}(\lambda,\mu)=\left(\beta^{(m)}(\lambda,\nu)+ 
\tilde\beta^{(m)}(\lambda) P(\nu)\right)*\alpha^{(1)}(\nu,\tilde\nu)\star
(T^{(1)})^{-1}(\tilde\nu,\mu),\\\nonumber
&&
m_i=1,\dots,D_i,\;\;i=1,2.
\end{eqnarray}

Similarly, to derive constraints assotiated with definition (\ref{S_A_bxi}), we apply operator  $* \alpha^{(m)}$ to the eq.(\ref{S_chi}) from the right. One obtains
\begin{eqnarray}\label{xi_check}
\bar \xi^{(m)} = \alpha^{(1)}\star (T^{(1)})^{-1}\star  \xi^{(1)} *\alpha^{(m)} -\bar\xi^{(1)}\star (T^{(1)})^{-1}\star \beta^{(1)}*\alpha^{(m)}.
\end{eqnarray}
Substitute eq.(\ref{S_beta_xi1}) for $\bar\xi^{(m)}$ and require that resulting equation is identity for any $\bar \xi^{(1)}$.  Then eq.(\ref{xi_check}) becomes decomposed into two following constraints:
\begin{eqnarray}\label{xi_0}
&&
\xi_0(\lambda,\nu)*\alpha^{(1)}(\nu,\mu)=0,
\\\label{S_bar_eta_m}
&&
\bar\eta^{(m)}(\lambda,\mu)=-(\bar T^{(1)})^{-1}(\lambda,\nu)\star \beta^{(1)}(\nu,\tilde\nu)*\alpha^{(m)}(\tilde\nu,\mu).
\end{eqnarray}
Eqs.(\ref{bxi_00}) and (\ref{xi_0}) represent two constraints for $\alpha^{(1)}$, $\beta^{(m)}$ and $\tilde\beta^{(m)}$. Eqs.(\ref{S_eta_m}) and (\ref{S_bar_eta_m}) with $m_1+m_2>2$ may be considered as definitions of $\eta^{(m)}$ and $\bar\eta^{(m)}$.
However, since 
 $\eta^{(1)}(\lambda,\mu)=\bar \eta^{(1)}(\lambda,\mu)={\cal{I}}_2(\lambda,\mu)$, 
eq.(\ref{S_bar_eta_m}) with $m_1=m_2=1$  yields:
\begin{eqnarray}
\label{S_bar_eta_1}
&&
\bar T^{(1)}(\lambda,\mu)=-\beta^{(1)}(\lambda,\nu)*\alpha^{(1)}(\nu,\mu),
\end{eqnarray}
In turn, eq.(\ref{S_eta_m}) with $m_1=m_2=1$ in view of eq.(\ref{S_bar_eta_1}) yields
\begin{eqnarray}
\label{S_eta_1}
&&
T^{(1)}(\lambda,\mu)=-\bar T^{(1)}(\lambda,\mu)+\tilde\beta^{(1)}(\lambda)
 P(\nu)*\alpha^{(1)}(\nu,\mu),
\end{eqnarray}
Both eqs.(\ref{S_bar_eta_1}) and  (\ref{S_eta_1}) may be treated as  constraints 
for $T^{(1)}$, $\bar T^{(1)}$, $\alpha^{(1)}$ and $\beta^{(1)}$. 

Now we 
simplify eq.(\ref{S_bar_eta_m}), $m_1+m_2>2$,  replacing $\beta^{(1)}\star \alpha^{(1)}$  in accordance with eq.(\ref{S_bar_eta_1}).
One gets
\begin{eqnarray}\label{S_bar_eta_m_mod}
&&
\bar \eta^{(m)}\star\eta^{(m)} =T^{(m)}\star(T^{(1)})^{-1} \;\;\stackrel{{\mbox{eq.(\ref{S_A_alp_bet_mod_gamma})}}}{\Rightarrow} \\\label{S_bar_eta_m_mod2}
&&
\gamma^{(m)}(\lambda,\mu)=(\bar T^{(1)})^{-1}(\lambda,\nu)\star  \bar T^{(m)}(\nu,\mu),\;\;m_1+m_2>2,
\end{eqnarray} 
which is the definition of $\gamma^{(m)}$. Deriving eq.(\ref{S_bar_eta_m_mod2}) we assume invertibility of the operator $\star T^{(m)}$, $\forall\;m$.

Constraints (\ref{bxi_00}) and (\ref{S_eta_m}) may be simplified   multiplying them  by $\bar \eta^{(m)}$ from the left, using definitions of $\alpha^{(m)}$ (\ref{S_alpha_m}), $\beta^{(m)}$ (\ref{S_beta_m}),  $\tilde \Gamma^{(m)}$ (\ref{S_A_alp_bet_mod2_tGamma}) and eliminating $\bar \eta^{(m)}\star \eta^{(m)}$  with eq.(\ref{S_bar_eta_m_mod}).
We obtain in result:
\begin{eqnarray}\label{bxi_0}\label{bar_xi_0}
&&
\Big((\bar T^{(1)})^{-1}(\lambda,\nu)\star \bar T^{(m)}(\nu,\tilde\nu)\star \beta^{(1)}(\tilde\nu,\tilde\mu)+
\tilde \Gamma^{(m)}(\lambda) P(\tilde\mu)\Big)*\bar \xi_0(\tilde\mu,\mu)=0,\;\;\forall \; m
,\\
\label{S_eta_m_2}
&&
\bar \eta^{(m)} \star\tilde \beta^{(m)} P*\alpha^{(1)} =T^{(m)} +\bar T^{(m)}\;\;\stackrel{{\mbox{eq.(\ref{S_A_alp_bet_mod2_tGamma})}}}{\Rightarrow} \\
\label{S_eta_m_22}
&&
\tilde\Gamma^{(m)}(\lambda)P(\nu)*\alpha^{(1)}(\nu,\mu)= T^{(m)}(\lambda,\mu)+\bar T^{(m)}(\lambda,\mu),\;\;m_1+m_2>2.
\end{eqnarray}
One has to take into account that $T^{(m)}$, $\bar T^{(m)}$ are represented by  eqs.(\ref{com_T_bT}) in terms of $T(\lambda,\mu)$, $\bar T(\lambda,\mu)$, $\tau^{(m)}(\lambda,\mu)$ and  $\bar \tau^{(m)}(\lambda,\mu)$.
Thus, we have obtained a set of constraints for the functions $\alpha^{(1)}$, $\beta^{(1)}$, $\tilde\Gamma^{(m)}$,  $T$, $\bar T$, $\tau^{(m)}$ and $\bar \tau^{(m)}$: 
eqs.(\ref{xi_0},\ref{S_bar_eta_1},\ref{S_eta_1},\ref{bxi_0},\ref{S_eta_m_22}). 

Note that, similar to  Sec.\ref{Section:C_internal}, 
all these constraints are generated  by   system (\ref{S_t}) and its compatibility condition. For this reasong we refer to them as the internal constraints in order to defer them from so-called external constraints wich will be introduced "by hand" for the purpose of derivation of the nonlinear PDEs, see Theorems 3.3, 3,4, 3,5. 

\subsubsection{System of compatible linear equations for $W(\lambda;t)$.}
{\bf Theorem 3.3}.
Let matrices $A^{(m)}(\lambda,\mu)$ satisfy  the following external constraint:
\begin{eqnarray}\label{S_constrain0}
\sum_{m_1=1}^{D_1} L^{(m_1)} P(\lambda)*A^{(m)}(\lambda,\mu) =  S^{(m_2)}P(\lambda)
,
\end{eqnarray}
where $L^{(m_1)}$ and $S^{(m_2)}$ are some $n_0\times n_0$  constant matrices.
Then
matrix function
$W(\lambda;t)$ obtained as a solution to the integral equation (\ref{Psi}) with $\chi$ defined by 
eq.(\ref{S_t}) is a solution to the following system of compatible linear equations
\begin{eqnarray}\label{S_lin}
&&
E^{(m_2)}(\lambda;t):=\\\nonumber
&&
\sum_{m_1=1}^{D_{1}}L^{(m_1)}\left( W_{t_m}(\lambda;t) + V^{(m)}(t) W (\lambda;t) 
+W(\mu;t) *A^{(m)}(\mu,\lambda;t) \right)= S^{(m_2)}W(\lambda;t),\\\nonumber
&&
m_i=1,\dots,D_i,\;\;i=1,2
\end{eqnarray}
where $V^{(m)}$ is given by eq.(\ref{V}).

{\bf Proof:} 
To derive eq.(\ref{S_lin}), we 
differentiate eq.(\ref{Psi}) with respect to $t_m$.  Then, in view of eq.(\ref{S_t}), 
one gets the following integral  equation:
\begin{eqnarray}\label{S_lin_1_proof}
&& 
{\cal{E}}^{(m)}(\mu;t):=P(\nu)*A^{(m)}(\nu,\lambda)*\chi(\lambda,\mu;t)=
\tilde E^{(m)}(\nu;t)*(\chi(\nu,\mu;t)+{\cal{I}}_1(\nu,\mu)) ,\\\nonumber
&&
\tilde E^{(m)}(\mu;t)=W_{t_m}(\mu;t) + V^{(m)}(t) W (\mu;t)
+W(\nu;t)* A^{(m)}(\nu,\mu).
\end{eqnarray}
Consider the following combination of eqs.(\ref{S_lin_1_proof}):
$\sum_{m_1=1}^{D_1} L^{(m_1)} {\cal{E}}^{m}$. Then, using constraint (\ref{S_constrain0}), one gets in result:
\begin{eqnarray}\label{S_lin_2}
&& \sum_{m_1=1}^{D_1} L^{(m_1)} {\cal{E}}^{m}:=
\sum_{m_1=1}^{D_1} L^{(m_1)} \tilde E^{(m)}(\nu;t)*(\chi(\nu,\mu;t)+{\cal{I}}_1(\nu,\mu)) =
\\\nonumber
&&
S^{(m_2)} W(\nu;t)*(\chi(\nu,\mu;t)+{\cal{I}}_1(\nu,\mu)) .
\end{eqnarray}
Since  operator $*(\chi(\nu,\mu;t)+{\cal{I}}_1(\nu,\mu))$ is invertable, eq.(\ref{S_lin_2}) is equivalent to  eq.(\ref{S_lin}).
$\blacksquare$

{\it Remark:} Similar to eq.(\ref{lin}), eq.(\ref{S_lin}) is, strictly speaking, nonlinear equation for $W(\lambda;t)$ since $V^{(m)}(t)$ are defined by eq.(\ref{V}) in terms of $W(\lambda;t)$.

System (\ref{S_lin}) is an analogy of the overdetermined system of linear equations in the classical 
inverse spectral transform method. According to this method, nonlinear PDEs for potentials of overdetermined linear system appear as compatibility conditions for this system. However, nonlinear PDEs may not be obtained by this method in our case because of the last term in the LHS of eq.(\ref{S_lin}). Instead of this, we represent another algorithm of derivation of the nonlinear PDEs in Secs.\ref{Section:S_first_order} and \ref{Section:second_order}.

\subsubsection{First order nonlinear PDEs for the fields $V^{(m)}(t)$, $m_i=1,\dots,D_i$, $i=1,2$.}
\label{Section:S_first_order}
{\bf Theorem 3.4.} In addition to eqs.(\ref{Psi},\ref{S_t}) and external constraint (\ref{S_constrain0}), we impose one more  external constraint:
\begin{eqnarray}\label{S_constrain1}
\sum_{m_2=1}^{D_{2}}A^{(m)}(\lambda,\nu) *\tilde A^{(n)}(\nu) R^{(m_2n)} =\sum_{j=1}^{D}
\tilde A^{(j)}(\lambda) P^{(m_1jn)},\;\;n_i=1,\dots,D_i,\;\;i=1,2,
\end{eqnarray}
where $R^{(m_2n)}$ and $P^{(m_1jn)}$ are some $n_0\times n_0$ constant matrices.
Then $n_0\times n_0$ matrix functions $V^{(m)}(t)$ are solutions to the following system of nonlinear PDEs:
\begin{eqnarray}\label{S_Q}
&&
\sum_{m=1}^{D}L^{(m_1)}\left( V^{(n)}_{t_m} + V^{(m)} V^{(n)} +V^{(m)}{\cal{A}}^{(n)}
\right)R^{(m_2n)}+\sum_{m_1=1}^{D_1}L^{(m_1)}\sum_{j=1}^{D} V^{(j)} P^{(m_1jn)}=\\\nonumber
&&
\sum_{m_2=1}^{D_2}
 S^{(m_2)}V^{(n)}R^{(m_2n)},\;\;{\cal{A}}^{(n)}=P(\lambda,\nu)* \tilde A^{(n)}(\nu),\;\;
n_i=1,\dots,D_i,\;\;i=1,2.
\end{eqnarray}

{\bf Proof:} 
Applying operator  $*\tilde A^{(n)}$ to eq.(\ref{S_lin}) from the right one gets the following 
equation
\begin{eqnarray}\label{S_2U}
&&
E^{(m_2n)}(t)=E^{(m_2)}(\lambda;t)* \tilde A^{(n)}(\lambda):=\\\nonumber
&&
\sum_{m_1=1}^{m_{t1}}L^{(m_1)}\left( V^{(n)}_{t_m} + V^{(m)} V^{(n)} +V^{(m)}{\cal{A}}^{(n)}
+U^{(mn)} \right)=
\sum_{m_2=1}^{m_{t2}} S^{(m_2)}V^{(n)},
\end{eqnarray}
which introduces a new set of fields $U^{(mn)}$, see eq.(\ref{UW}).
Due to the constraint (\ref{S_constrain1}), we may eliminate these fields using a 
proper combinations of eqs.(\ref{S_2U}). Namely, combinations $\sum_{m_2=1}^{D_2} E^{(m_2n)} R^{(m_2n)}$ results in the system 
(\ref{S_Q}). $\blacksquare$

\paragraph{Reduction 1} The derived equation (\ref{S_Q}) admits the following reduction for its coefficients: 
\begin{eqnarray}\label{part_constr}
 P^{(m_1jn)}=-{\cal{A}}^{(n)}R^{(j_2n)} \delta_{j_1m_1},\;\;S^{(m_2)}=0.
\end{eqnarray}
Then  eq.(\ref{S_Q}) reads:
\begin{eqnarray}\label{S_Q_simple}
&&
\sum_{m=1}^{D}L^{(m_1)}\left(  V^{(n)}_{t_m} + V^{(m)}  V^{(n)}\right) R^{(m_2n)}=0,
\end{eqnarray}
which is represented in the introduction, see eq.(\ref{nl1_intr}).
This reduction does not effect constraint  (\ref{S_constrain0}) while constraint (\ref{S_constrain1})
reads:
\begin{eqnarray}
\label{S_constrain1_red1}
\sum_{m_2=1}^{D_{2}}(A^{(m)}(\lambda,\nu)+\tilde A^{(m)}(\lambda) P(\nu))* \tilde A^{(n)}(\nu) R^{(m_2n)} =0,\;\;n_i=1,\dots,D_i,\;\;i=1,2.
\end{eqnarray}

\paragraph{Reduction 2}
Along with reduction (\ref{part_constr}), we 
consider reduction (\ref{red2},\ref{red22}) with
\begin{eqnarray}\label{AC}
&&
C^{(n)}R^{(m_2n)}=R^{(m_2)}.
\end{eqnarray}
where  $C^{(m)}$ and $R^{(m_2)}$ are some $n_0\times n_0$   constant matrices.
Then the system (\ref{S_Q_simple}) reduces to the single PDE:
\begin{eqnarray}\label{S_Q_simple_red2}
&&
\sum_{m=1}^{D}L^{(m_1)}\left(  V_{t_m} + V C^{(m)}  V\right) R^{(m_2)}=0,
\end{eqnarray}
which is written in the Introduction, see eq.(\ref{nl1_intr}).
Reduction (\ref{AC}) does not effect constraint  (\ref{S_constrain0}), while constraint  (\ref{S_constrain1_red1})
reduces to the following single equation
\begin{eqnarray}
\label{S_constrain1_red2}
\sum_{m_2=1}^{D_{2}}(A^{(m)}(\lambda,\nu)+\tilde A(\lambda) C^{(m)} P(\nu)) *\tilde A(\nu) R^{(m_2)} =0.
\end{eqnarray}
If, in addition, $C^{(m_1m_2)}$, $L^{(m_1)}$ and $R^{(m_2)}$ are diagonal, 
\begin{eqnarray}\label{AC2}
&&
C^{(m_1m_2)}=-C^{(m_2m_1)}, \;\;R^{(m_2)} \equiv L^{(m_2)},\;\;V_{t_{m_1m_2}}=-V_{t_{m_2m_1}},
\;\;D_1=D_2=D_0,
\end{eqnarray}
then nonlinear equation (\ref{S_Q_simple_red2}) reduces to the following equation
\begin{eqnarray}\label{S_Q_simple_red22}
&&
\sum_{{m_1,m_2=1}\atop{m_2>m_1}}^{D_{0}}\left( L^{(m_1)} V_{t_{m_1m_2}}L^{(m_2)}- 
L^{(m_2)} V_{t_{m_1m_2}}L^{(m_1)}+\right.\\\nonumber
&&\left.
 L^{(m_1)}VC^{(m_1m_2)}VL^{(m_2)}  - L^{(m_2)}VC^{(m_1m_2)})  VL^{(m_1)}\right) =0,
\end{eqnarray}
which is a multidimensional generalization of the classical (2+1)-dimensional $S$-integrable $N$-wave equation (\ref{S_Nwave}).

Eq.(\ref{S_Q_simple_red22}) admits reduction
\begin{eqnarray}
t_{m_1m_2}=- i \tau_{m_1m_2},\;\;\;V=-V^{+},\;\;m_2>m_1,
\end{eqnarray}
see eq.(\ref{S_Q_simple_red_intr}),
which is important for physical applications.

\subsubsection{Second order nonlinear PDEs for $V^{(m)}(t)$, $m_i=1,\dots,D_i$, $i=1,2$}
\label{Section:second_order}
{\bf Theorem 3.5.} 
In addition to eqs.(\ref{Psi},\ref{S_t}) and external constraint (\ref{S_constrain0}) we impose one more external constraint on the matrix functions   $A^{(m)}(\lambda,\nu)$ and $\tilde A^{(m)}(\lambda)$ (instead of constraint (\ref{S_constrain1})):
\begin{eqnarray}\label{S_constrain2}
&&
\sum_{n=1}^D\sum_{m_2=1}^{D_2}A^{(m)}(\lambda,\nu)*A^{(n)}(\nu,\mu) *\tilde A^{(l)}(\mu) R^{(m_2nl)} =\\\nonumber
&&
\sum_{n,p=1}^{D}A^{(n)}(\lambda,\mu)*\tilde A^{(p)}(\mu) P^{(m_1npl)}+
\sum_{n=1}^{D}\tilde A^{(n)}(\lambda) P^{(m_1nl)},
\end{eqnarray}
where 
$R^{(m_2nl)}$, $P^{(m_1npl)}$, and $P^{(m_1nl)}$ are some $n_0\times n_0$  constant matrices. 
Let $S^{(m_2)}=0$ for the sake of simplicity. 
Then $n_0\times n_0$ matrix functions $V^{(m)}(t)$
are solutions to the following system of nonlinear PDEs:
\begin{eqnarray}\label{S_2Q_V}
&&
\sum_{m,n=1}^{D} L^{(m_1)}\left(
U^{(nl)}_{t_m} +V^{(m)}U^{(nl)}+V^{(m)} {\cal{A}}^{(nl)}
\right)R^{(m_2nl)}
+\\\nonumber
&&
\sum_{m_1=1}^{D_1} L^{(m_1)}\left(\sum_{n,p=1}^D U^{(np)} P^{(m_1npl)}+
\sum_{n=1}^D V^{(n)} P^{(m_1nl)}\right)
=0,\\\nonumber
&&
 {\cal{A}}^{(nl)}= P(\lambda) *A^{(n)}(\lambda,\mu)*\tilde A^{(l)}(\mu),\;\;l_i=1,\dots,D_i,\;\;i=1,2,
\end{eqnarray}
where fields $U^{(mn)}$ are related with $V^{(m)}$  due to eq.(\ref{S_2U}):
\begin{eqnarray}\label{S_2U2}
&&
\sum_{m_1=1}^{D_1}L^{(m_1)}\left( V^{(n)}_{t_m} + V^{(m)} V^{(n)} +V^{(m)}{\cal{A}}^{(n)}
+U^{(mn)} \right)=0,\\\nonumber
&&
{\cal{A}}^{(n)}=P(\lambda) *\tilde A^{(n)}(\lambda).
\end{eqnarray}

{\bf Proof:} First of all, applying operator $*\tilde A^{(n)} $  to eq.(\ref{S_lin})  from the right one gets 
equation (\ref{S_2U2}),
which introduces a new set of fields $U^{(mn)}(t)$, see eq.(\ref{S_UW}). 
This step is equivalent to the first step in  derivation of eq.(\ref{S_Q}).
However, we may not eliminate these fields from the system (\ref{S_2U2}) because constraint (\ref{S_constrain1}) is not valid in this Theorem.
Instead of this, we 
 derive nonlinear equations for  fields $U^{(mn)}(t)$
applying operator $*A^{(n)}*\tilde A^{(l)}$ to  eq.(\ref{S_lin})  from the right. 
One gets
\begin{eqnarray}\label{S_2Q0}
&&
E^{(m_2nl)}(t)=E^{(m_2)}(\lambda;t)*A^{(n)}(\lambda,\mu)*\tilde A^{(l)}(\mu) :=
\\\nonumber
&&
\sum_{m_1=1}^{D_1} L^{(m_1)}\left(
U^{(nl)}_{t_m} +V^{(m)}U^{(nl)}+V^{(m)} {\cal{A}}^{(nl)}+
U^{(mnl)}\right)
=0,
\end{eqnarray}
where  the fields $U^{(mnl)}$  are defined by eq.(\ref{U_mnl}).
Due to the constraint (\ref{S_constrain2}), 
these fields may be eliminated taking a proper 
combination of equations (\ref{S_2Q0}), namely,
$\sum_{m_2=1}^{D_2} E^{(m_2nl)} R^{(m_2nl)}$. In result, one obtains 
eq.(\ref{S_2Q_V}). $\blacksquare$

\paragraph{Reduction 1.}
Let
\begin{eqnarray}\label{S_red12}
&&
P^{(m_1npl)}=\delta_{m_1n_1} \tilde P^{(n_2pl)},\;\;
P^{(m_1nl)}=\delta_{m_1n_1} \tilde P^{(n_2l)},\\\nonumber
&&
 \tilde P^{(n_2l)}=
\sum_{p=1}^D({\cal{A}}^{(p)} \tilde P^{(n_2pl)} - {\cal{A}}^{(pl)} R^{(n_2 pl)}).
\end{eqnarray}
Then eq.(\ref{S_2Q_V}) reduces to the following one:
\begin{eqnarray}\label{S_2Q_V_red1}
&&
\sum_{m,n=1}^D L^{(m_1)}\left[\left(
U^{(nl)}_{t_m} + V^{(m)} U^{(nl)}
\right) R^{(m_2nl)} +\left(
V^{(n)}_{t_m}+ V^{(m)} V^{(n)}
\right) \tilde P^{(m_2 nl)}\right]=0,\\\nonumber
&&
l_i=1,\dots,D_i,\;\;i=1,2.
\end{eqnarray}
Constraint (\ref{S_constrain0}) remains the same (with $S^{(m_2)}=0$) while constraint 
(\ref{S_constrain2}) reads
\begin{eqnarray}\nonumber
&&
\sum_{n=1}^D\sum_{m_2=1}^{D_2}\left(A^{(m)}(\lambda,\nu)+\tilde A^{(m)}(\lambda) P(\nu)\right)*\left(A^{(n)}(\nu,\mu)* \tilde A^{(l)}(\mu) R^{(m_2nl)}-
\tilde A^{(n)}(\tilde\nu)\tilde P^{(m_2 nl)}\right),\\\label{S_constrain2_red1}
&&
m_1=1,\dots,D_1,\;\;l_i=1,\dots,D_i,\;\;i=1,2.
\end{eqnarray}

\paragraph{Reduction 2.}
Along with reduction (\ref{S_red12}) we 
consider reduction (\ref{red2},\ref{red22}) together with the following conditions
\begin{eqnarray}
U^{(ml)}(t)=U^{(m)}(t) C^{(l)},\;\;C^{(l)}R^{(m_2 nl)}=R^{(m_2 n)} ,\;\;C^{(n)}\tilde P^{(m_2 nl)}=P^{(m_2 n)}.
\end{eqnarray}
Then the system (\ref{S_2Q_V_red1}) reduces to the following single PDE:
\begin{eqnarray}\label{S_2Q_V_red2}
&&
\sum_{m,n=1}^D L^{(m_1)}\left[\left(
U^{(n)}_{t_m} + V C^{(m)} U^{(n)}
\right) R^{(m_2n)} +\left(
V_{t_m}+ VC^{(m)} V
\right) P^{(m_2 n)}\right]=0
\end{eqnarray}
This reduction does not effect constraint (\ref{S_constrain0}), while constraint (\ref{S_constrain2_red1})
reduces to the following single equation:
\begin{eqnarray}\label{S_constrain2_red2}
\sum_{m_2=1}^{D_2}\sum_{n=1}^D\left(A^{(m)}(\lambda,\nu)+\tilde A(\lambda) C^{(m)} P(\nu)\right)*\left(A^{(n)}(\nu,\mu)*
 \tilde A(\mu) R^{(m_2n)}-
\tilde A(\nu) P^{(m_2 n)}\right).
\end{eqnarray}

Emphasize that systems (\ref{S_Q}) and (\ref{S_2Q_V},\ref{S_2U2}) do not represent 
commuting flows since constraints (\ref{S_constrain1}) and (\ref{S_constrain2}) are not compatible in general.

\subsubsection{Solutions to the first-order 
equation (\ref{S_Q_simple_red2})}

\paragraph{On the dimensionality of the available solution space.}
The important question is whether the derived nonlinear DPEs are completely integrable. In other words, regarding eq.(\ref{S_Q_simple_red2}), is it possible to introduce a proper number of arbitrary functions of $D_1D_2-1$ variables in the solution space.

In order to clarify this problem,  first of all, let us 
consider small $\chi$. Then, using equation (\ref{Psi}), we approximate $V$ by eq.(\ref{V_approx}). 
If  $\tau^{(m)}(\nu)$ and $\bar \tau^{(m)}(\nu)$ are arbitrary functions of arguments, 
then the above expression for $V$  involves two arbitrary functions  of $D_1D_2$ variables $t_m$, $m_i=1,\dots,D_i$, $i=1,2$: 
\begin{eqnarray}\label{S_F}
&&
F_1(t)=P*\alpha^{(1)}\star (T^{(1)})^{-1}\star \xi^{(1)} * \alpha^{(1)}\star \tilde \Gamma=
\int g_{11}(\nu) e^{\sum_{i=1}^D\tau^{(i)} t_i} g_{12}(\nu) d\Omega_2(\nu),\\\nonumber
&&
F_2(t)=P*\alpha^{(1)}*\bar \xi^{(1)}\star (\bar T^{(1)})^{-1}\star \beta^{(1)}*\alpha^{(1)}\star \tilde \Gamma=
\int g_{21}(\nu) e^{\sum_{i=1}^D\tau^{(i)} t_i} g_{22}(\nu) d\Omega_2(\nu),
\end{eqnarray}
where
\begin{eqnarray}
&&
g_{11}=P*\alpha^{(1)}\star (T^{(1)})^{-1}\star T,\;\;g_{12}=T^{-1}\star \xi_0* \alpha^{(1)}\star \tilde \Gamma,\\\nonumber
&&
g_{21}=P*\alpha^{(1)}\star \bar\xi_0\star \bar T^{-1},\;\;
g_{22}=\bar T\star (\bar T^{(1)})^{-1}\star \beta^{(1)}*\alpha^{(1)}\star \tilde \Gamma.
\end{eqnarray}
However, not all $\tau^{(m)}(\nu)$ and $\bar \tau^{(m)}(\nu)$ are arbitrary since we have to resolve constraints\newline
(\ref{xi_0},\ref{S_bar_eta_1},\ref{S_eta_1},\ref{bxi_0},\ref{S_eta_m_22},\ref{S_constrain0},\ref{S_constrain1_red2}). 
Constraints (\ref{xi_0},\ref{bxi_0}) may be satisfied using a special structures of  $\alpha^{(1)}$, $\beta^{(1)}$, $P$, $\bar \xi_0$, $\xi_0$ 
(see example of explicite solution, eqs.(\ref{structure})), which does not reduce the dimensionality of the solution space. Constraints (\ref{S_bar_eta_1}) may be considered as constraints for $\alpha^{(1)}$ and $\beta^{(1)}$.
Constraints (\ref{S_eta_1},\ref{S_eta_m_22})  relate
 $\bar T^{(m)}$ with $T^{(m)}$ and $\tilde \Gamma$, so that, generally speaking, only one of functions (\ref{S_F}) remains arbitrary. 
Two remaining constraints (\ref{S_constrain0},\ref{S_constrain1_red2}) introduce $D_2$ and $D_1$ relations among parameters  $\bar\tau^{(m)}$, $m_i=1,\dots,D_i$, $i=1,2$, which, in general, reduces significantly the dimensionality of the solution space. However, this depends on the particular choice of the coefficients $L^{(mj_1)}$, $R^{(m_2)}$ and $C^{(m)}$. 

This is a preliminary analysis which suggests us to look for examples of completely integrable PDEs in the derived class of new nonlinear PDEs.

\paragraph{Construction of explicite solutions.}
\label{S_Section:first_order}
We construct explicite solutions in the form of rational functions of exponents for the first order nonlinear PDE 
(\ref{S_Q_simple_red2}) with $d\Omega_i(\nu)$, $i=1,2$, given by eqs.(\ref{Omega}). 
Along with notations (\ref{notations_discr}) we use the following ones:
\begin{eqnarray}\label{notations_discr2}
\hat{\bar\xi}_0=\left[
\begin{array}{ccc}
\bar \xi_0(a_1,b_1)&\cdots& \bar \xi_0(a_1,b_N)\cr
\cdots&\cdots&\cdots\cr
\bar \xi_0(a_M,b_1)&\cdots& \bar \xi_0(a_M,b_N)
\end{array}
\right],\;\;\hat{\bar\tau}^{(m)}={\mbox{diag}}(\tau^{(m)}(b_1),\cdots,\tau^{(m)}(b_N)),\\\nonumber
\;\;T(\lambda,\mu)=\bar T(\lambda,\mu)={\cal{I}}_2(\lambda,\mu).
\end{eqnarray}
Field $V(t)$ is represented  by eq.(\ref{V_sol}) where 
function $\chi$ is given by eq.(\ref{S_chi}). In turn, the functions $\xi^{(1)}$ and $\bar\xi^{(1)}$   
are  given by eqs.(\ref{S_xi_t}) and (\ref{S_bar_xi_t}). In result, using notations (\ref{notations_discr},\ref{notations_discr2}), we obtain the following formula for $\hat \chi$:
\begin{eqnarray}\label{S_chi_sol}
\hat \chi=\hat \alpha^{(1)}(\hat \tau^{(1)})^{-1} e^{\sum_{i=1}^{D} \hat \tau^{(i)} t_i} \hat \xi^{(0)}+\hat {\bar\xi}^{(0)}e^{\sum_{i=1}^{D} \hat{\bar\tau}^{(i)} t_i}(\hat{\bar\tau}^{(1)})^{-1}\hat\beta^{(1)}.
\end{eqnarray}
Finally, constraints (\ref{xi_0},\ref{S_bar_eta_1},\ref{S_eta_1},\ref{bxi_0},\ref{S_eta_m_22},\ref{S_constrain0},\ref{S_constrain1_red2})  must be satisfied,
which read in view of notations (\ref{notations_discr},\ref{notations_discr2}) as follows, $m_i=1,\dots,D_i$, $i=1,2$:
\begin{eqnarray}
\label{xi_0_expl}
&&
\hat \xi_0\hat \alpha^{(1)}=0
,\\
\label{S_bar_eta_1_expl}
&&
\hat{\bar\tau}^{(1)}=-\hat\beta^{(1)}\hat \alpha^{(1)},
\\\label{S_eta_1_expl}
\label{S_eta_m_22_expl}\label{S_eta_m_2_red}
&&
\hat{\tilde\Gamma} C^{(m)}\hat P\hat \alpha^{(1)}=\hat \tau^{(m)}+\hat{\bar\tau}^{(m)}\\
\label{bxi_0_expl}
&&\Big((\hat{\bar \tau}^{(1)})^{-1} \hat{\bar \tau}^{(m)} \hat\beta^{(1)}+\hat{\tilde \Gamma} C^{(m)} \hat P\Big)
\hat{\bar \xi}_0=0
,\\
\label{S_constrain0_expl}\label{S_constrain01}
&&
\sum_{m_1=1}^{D_1} L^{(m_1)} \hat P\hat \alpha^{(1)}(\hat{\bar\tau}^{(1)})^{-1}
\hat{\bar\tau}^{(m)}\hat \beta^{(1)} =  0
,
\\
\label{S_constrain1_red2_expl}
&&
\sum_{m_2=1}^{D_2}\hat \alpha^{(1)}((\hat{\bar\tau}^{(1)})^{-1}
\hat{\bar\tau}^{(m)}\beta^{(1)}+\hat{\tilde \Gamma} C^{(m)}\hat P) \hat \alpha^{(1)}
\hat{\tilde \Gamma} R^{(m_2)} =0.
\end{eqnarray}
Here we combine equations (\ref{S_eta_1}) and (\ref{S_eta_m_22}) into the single equation (\ref{S_eta_m_2_red}).
Let us transform some of eqs.(\ref{xi_0_expl}-\ref{S_constrain1_red2_expl}). For instance,  
considering only particular solutions to eq.(\ref{bxi_0_expl}) we assume
\begin{eqnarray}\label{xi_0_a2}
\hat \beta^{(1)} \hat{\bar \xi}_0 = \hat P\hat{\bar \xi}_0 =0.
\end{eqnarray}
In other words, the rows of $\hat \beta^{(1)}$ and $\hat P$ are orthogonal to the columns of 
$\hat{\bar \xi}_0$. Similarly, eq.(\ref{xi_0_expl})  means that the rows of 
$\hat \xi_0$ are orthogonal to the columns  of $\hat \alpha^{(1)}$. 

In turn, constraint (\ref{S_constrain1_red2_expl})
may be transformed into the following one using eq.(\ref{S_bar_eta_1_expl}). 
\begin{eqnarray}\label{S_constrain12}
&&
\sum_{m_2=1}^{D_2}\hat \alpha^{(1)} (\hat{\bar\tau}^{(m)} -\hat{\tilde\Gamma} 
C^{(m)}\hat P\hat \alpha^{(1)} )\hat{\tilde\Gamma}
R^{(m_2)} =
0
\end{eqnarray}
Since $\hat{\bar\tau}^{(1)}$ is invertable, one requires  $M>N$, which is evident due to eq.(\ref{S_bar_eta_1_expl}). 
Then we may rewrite eq.(\ref{S_constrain12}) 
without factor $\hat\alpha^{(1)}$ as follows:
\begin{eqnarray}\label{S_constrain13_part}
&&
\sum_{m_2=1}^{D_{2}}\left(\hat{\bar\tau}^{(m)} -
\hat{\tilde\Gamma} C^{(m)}\hat P\alpha^{(1)}\right)\hat{\tilde\Gamma}
R^{(m_2)} = 0.
\end{eqnarray}

All in all, the following constraints must be satisfied:  (\ref{xi_0_expl}- \ref{S_eta_1_expl},
\ref{S_constrain01},\ref{xi_0_a2},\ref{S_constrain13_part}).

\paragraph{Simple examples of explicite solutions.}

We obtain a particular solution to the four-dimensional nonlinear PDE (\ref{S_Q_simple_red2}), i.e. $D_1=D_2=2$. Remember that we use double indices so that, for instance, $\hat{\tilde \beta}^{(m)}=\hat{\tilde\beta}^{(m_1m_2)}$.
Let $M=5$, $N=2$, $n_0=2$,
$L^{(1)}=R^{(1)}=I$, $L^{(2)}={\mbox{diag}}(l_1,l_2)$, $R^{(2)}={\mbox{diag}}(r_1,r_2)$,
 $C^{(n_1n_2)}={\mbox{diag}}(c^{(n_1n_2)}_1,c^{(n_1n_2)}_2)$, $c^{(11)}_1=c^{(11)}_2=1$.
To satisfy eqs.(\ref{xi_0_expl}) and (\ref{xi_0_a2}) we use the following structures of matrices:
\begin{eqnarray}\label{structure}
&&
\hat \beta^{(11)}= \left[
\begin{array}{ccccc}
Z&Z&I&Z&J_1\cr
Z&Z&Z&J_1&I
\end{array}
\right],\;\;
\hat P=\left[
\begin{array}{cccccccccc}
0&0&0&0&-1&0&0&1&-1&1\cr
0&0&0&0&-2&1&-1&2&1&2
\end{array}
\right],\\\nonumber
&&\hat{\tilde\Gamma}^{(11)}=\left[
\begin{array}{c}
J_0\cr
Z
\end{array}
\right],\;\;
\hat \xi_0=\left[
\begin{array}{ccccc}
J_0&Z&Z&Z&Z\cr
Z&-J_2&Z&Z&Z
\end{array}
\right],\;\;\hat{\bar\xi}_0=\left[
\begin{array}{cc}
K_2&Z\cr
Z &J_0\cr
Z&Z\cr
Z&Z\cr
Z&Z
\end{array}
\right],\;\;
\hat\alpha^{(11)}=\left[
\begin{array}{c}
Z_4\cr
K_{6,4}
\end{array}
\right],
\end{eqnarray}
where $Z$ and $I$ are  $2\times 2$ zero and  identity matrices respectively,
$Z_4$ is $4\times 4$ zero matrix, $K_{6,4}$ is $6\times 4$ constant 
matrix which will be defined below,
\begin{eqnarray}
J_0={\mbox{diag}}(1,-1),\;\;\;
J_1=
\left[
\begin{array}{cc}
0&1\cr
1&0
\end{array}
\right],\;\;\;J_2=
\left[
\begin{array}{cc}
0&1\cr
-1&0
\end{array}
\right],\;\;K_2=
\left[
\begin{array}{cc}
s_3&1\cr
s_4&s_5
\end{array}
\right]
\end{eqnarray}
We take $\hat{\bar\tau}^{(11)}={\mbox{diag}}(1,2,3,4)$.

To satisfy 
eqs.(\ref{S_bar_eta_1_expl}) and (\ref{S_eta_1_expl}) ($m_1=m_2=1$) we take
\begin{eqnarray}
&&
K_{6,4}=\left[
\begin{array}{cccc}\displaystyle
\frac{1}{5} (2 s_1 -3) & s_2 & -1&\displaystyle\frac{4}{3}\cr
s_1 & s_2-2 &\displaystyle\frac{1}{2} &\displaystyle\frac{4}{3}\cr
\displaystyle\frac{2}{5} (s_1+1) & s_2 & -1 &\displaystyle-\frac{8}{3}\cr
s_1 &s_2 & \displaystyle-\frac{5}{2} &\displaystyle\frac{4}{3}\cr
-s_1&-s_2&\displaystyle-\frac{1}{2}& \displaystyle-\frac{4}{3}\cr
\displaystyle-\frac{2}{5} (s_1+1) & -s_2 & 1 &\displaystyle-\frac{4}{3}
\end{array}
\right],\\\nonumber
&&
\tau^{(11)}={\mbox{diag}}\left(\frac{2}{5}(3 s_1-2,\;3 s_2, \;-3, \;-4\right).
\end{eqnarray}
To satisfy eqs.(\ref{S_constrain01}) and (\ref{S_constrain13_part}), $m_1+m_2>2$,   we take
matrices $\hat\Gamma^{(n_1n_2)}$ in the following  form:
\begin{eqnarray}
&&
\hat{\tilde\Gamma}^{(12)}=
\left[
\begin{array}{c}
J_3 \cr Z
\end{array}
\right]
,\;\;
\hat{\tilde\Gamma}^{(21)}=\left[
\begin{array}{c}
J_4 \cr 
Z
\end{array}
\right]
,\;\;
\hat{\tilde\Gamma}^{(22)}=\left[
\begin{array}{c}
J_5 \cr
Z
\end{array}
\right]
,\\\nonumber
&&
J_3={\mbox{diag}}\left(
-\frac{1}{r_1}(1+c^{(21)}_1 l_1 + c^{(22)}_1 l_1 r_1),
\frac{1}{r_2}(1+c^{(21)}_2 l_2 +c^{(22)}_2 l_2 r_2)
\right),
\\\nonumber
&&
J_4={\mbox{diag}}(c^{(21)}_1, -c^{(21)}_2),\;\;
J_5={\mbox{diag}}(c^{(22)}_1, -c^{(22)}_2).
\end{eqnarray}
Elements of matrix $C^{(12)}$ must be defined as follows:
\begin{eqnarray}\label{c12}
c^{(12)}_i=-\frac{1}{r_i}(1+l_i(c^{(21)}_i+c^{(22)}_i r_i)),\;\;i=1,2,
\end{eqnarray}
and $\hat\tau^{(n_1,n_2)}$, $\hat{\bar\tau}^{(n_1,n_2)}$
are following: 
\begin{eqnarray}
&&
\hat\tau^{(12)}={\mbox{diag}}\left(\frac{4-6s_1}{5 r_1},\;-\frac{3s_2}{r_2},\;\hat\tau^{(12)}_3,\;\hat\tau^{(12)}_4
\right),\\\nonumber
&&
\hat\tau^{(21)}={\mbox{diag}}\left(
\frac{1}{5}c^{(21)}_1(6s_1+1)+\frac{1}{l_1},\;c^{(21)}_2(2+3 s_2)+\frac{2}{l_2},\;\hat\tau^{(21)}_3,\;\hat\tau^{(21)}_4
\right)
,
\\\nonumber
&&
\hat\tau^{(22)}={\mbox{diag}}\left(
-\frac{5+c^{(21)}_1(1+6s_1)l_1}{5l_1r_1},\;-\frac{2+c^{(21)}_2(2+3s_2)l_2}{l_2 r_2},\;\hat\tau^{(22)}_3,\;\hat\tau^{(22)}_4
\right),
\\\nonumber
&&
\bar\tau^{(12)}={\mbox{diag}}\left(-\frac{1}{5r_1}(5 +(1+6 s_1)l_1(c^{(21)}_1 +
c^{(22)}_1 r_1)),\right.\;\\\nonumber
&&\left.
-\frac{1}{r_2}(2 +(2+3 s_2)l_2(c^{(21)}_2 +
c^{(22)}_2 r_2)),-\tau^{(12)}_3,-\tau^{(12)}_4\right),\\\nonumber
&&
\hat{\bar\tau}^{(21)}=-{\mbox{diag}}\left(
\frac{1}{l_1},\;\frac{2}{l_2},\;\tau^{(21)}_3,\;\tau^{(21)}_4
\right),\\\nonumber
&&
\hat{\bar\tau}^{(22)}={\mbox{diag}}\left(\frac{1}{5l_1r_1}(5 +(1+6 s_1)l_1(c^{(21)}_1 +
c^{(22)}_1 r_1)),\right.\;\\\nonumber
&&\left.
\frac{1}{l_2r_2}(2 +(2+3 s_2)l_2(c^{(21)}_2 +
c^{(22)}_2 r_2)),-\hat \tau^{(22)}_3,-\hat \tau^{(22)}_4\right),
\end{eqnarray}

Introduce positive parameters $p_i$, $i=1,2,3,4$, by the following formulas:
\begin{eqnarray}\label{ss1}
s_2=\frac{p_3}{18 s_1-12},\;\;s_3=-\frac{2}{5p_3}(3 s_1 -2) p_1,\\\nonumber
s_4=\frac{1}{5p_3}(p_3 p_4-p_1p_2),\;\;
s_5=\frac{p_2}{6 s_1-4},
\end{eqnarray}
so that solution $V(t)$ reads
\begin{eqnarray}\label{V11_1}
&&
V(t)=\frac{1}{D}
\left[
\begin{array}{cc}\displaystyle
f_{11}(p_2 e^{\hat\eta_2}+p_3 e^{\hat\eta_1+\hat\eta_2}) 
 & \displaystyle f_{12}e^{\hat\eta_3}\cr
\displaystyle f_{21} e^{\hat\eta_1+\hat\eta_2-\hat\eta_3} 
&
\displaystyle f_{22} ( p_1 e^{\hat\eta_1} + p_3 e^{\hat\eta_1+\hat\eta_2})
\end{array}
\right],\\\nonumber
&&
D=p_1 e^{\hat\eta_1} +
p_2 e^{\hat\eta_2}+p_3 e^{\hat\eta_1+\hat\eta_2}+p_4 ,\\\nonumber
&&
f_{11}=-\frac{1}{5}(1+6s_1),\;\;f_{12}=\frac{ p_3 (1+6s_1)}{2(3 s_1-2)},\\\nonumber
&&
f_{21}=\frac{1}{5 p_3}(p_1p_2-p_3p_4)(12 s_1 +p_3-8),\;\;f_{22}=
-\frac{12 s_1 +p_3-8}{2(3 s_1-2)}.
\end{eqnarray}
Here
\begin{eqnarray}
\hat\eta_1=\eta_1-\eta_3,\;\;\hat\eta_2=\eta_2-\eta_3,\;\;\hat\eta_3=\eta_4-\eta_3,\;\;
\eta_{i}=\sum_{m_1,m_2=1}^2 a^{(i)}_{m_1m_2} t_{m_1m_2},\;\;i=1,2,3,4,
\end{eqnarray}
\begin{eqnarray}
&&
a^{(1)}_{11}=\frac{1 + 6 s_1}{5},\;\;
a^{(1)}_{12}=\frac{1}{2 r_2}\left(4 + \frac{p_3}{3 s_1-2}\right),\\\nonumber
&&
a^{(1)}_{21}=\frac{1}{5 l_2}\left(10 +  l_2  c^{(21)}_1 (1 + 6 s_1)\right),\\\nonumber
&&
a^{(1)}_{22}=\frac{1}{5}\left(\frac{5 +  l_1  c^{(21)}_1}{ l_1 r_1} - \frac{10  c^{(21)}_2}{r_2} +  c^{(22)}_1 - 10  c^{(22)}_2 + 
  \frac{5 ( c^{(21)}_2 + r_2  c^{(22)}_2) p_3}{r_2 (4 - 6 s_1)} + 
  \frac{6 ( c^{(21)}_1 + r_1  c^{(22)}_1) s_1}{r_1}\right),
\end{eqnarray}
\begin{eqnarray}
&&
a^{(2)}_{11}=2 + \frac{p_3}{ 6 s_1-4},\\\nonumber
&&
a^{(2)}_{12}=\frac{1}{5}\left(\frac{(1 +  l_1  c^{(21)}_1) (1 + 6 s_1)}{r_1} +  l_1  c^{(22)}_1 (1 + 6 s_1) - 
  \frac{5  l_2 ( c^{(21)}_2 + r_2  c^{(22)}_2) (-8 + p_3 + 12 s_1)}{
   2 r_2 (3 s_1-2)}\right),\\\nonumber
&&
a^{(2)}_{21}=\frac{2}{ l_2} +  c^{(21)}_2 \left(2 + \frac{p_3}{ 6 s_1-4}\right),\;\;
a^{(2)}_{22}=\frac{5 +  l_1  c^{(21)}_1 (1 + 6 s_1)}{5  l_1 r_1} + 
  c^{(21)}_2 \left(-\frac{2}{r_2} + \frac{p_3}{4 r_2 - 6 r_2 s_1}\right)
,\end{eqnarray}
\begin{eqnarray}
&&
a^{(3)}_{11}=\frac{11}{5} + \frac{6 s_1}{5} + \frac{p_3}{ 6 s_1-4},\;\;
a^{(3)}_{12}=-\frac{ l_2 ( c^{(21)}_2 + r_2  c^{(22)}_2) (-8 + p_3 + 12 s_1)}{
 2 r_2 ( 3 s_1-2)},\\\nonumber
&&
a^{(3)}_{21}=\frac{2}{ l_2} + \frac{ c^{(21)}_1 (1 + 6 s_1)}{5} + 
  c^{(21)}_2 \left(2 + \frac{p_3}{ 6 s_1-4}\right),\\\nonumber
&&
a^{(3)}_{22}=\frac{-5  l_1 r_1  c^{(21)}_2 (-8 + p_3 + 12 s_1) + 
  2 r_2 ( 3 s_1-2) (5 +  l_1 ( c^{(21)}_1 + r_1  c^{(22)}_1) 
     (1 + 6 s_1))}{10  l_1 r_1 r_2 (3 s_1-2)}
,
\end{eqnarray}
\begin{eqnarray}
&&
a^{(4)}_{11}=\frac{6 (1 + s_1)}{5},\\\nonumber
&&
a^{(4)}_{12}=\frac{1}{5 r_1}(5 +  l_1 ( c^{(21)}_1 + r_1  c^{(22)}_1) (1 + 6 s_1)) + 
 \frac{p_3 -  l_2 ( c^{(21)}_2 + r_2  c^{(22)}_2) (-8 + p_3 + 12 s_1)}{2 r_2 (3 s_1-2)},\\\nonumber
&&
a^{(4)}_{21}= \frac{1}{l_1} + \frac{c^{(21)}_1 (1 + 6 s_1)}{5},\;\;
a^{(4)}_{22}=\frac{2}{l_2 r_2}.
\end{eqnarray}

Note that not all constant matrices may be arbitrary diagonal matrices in eq.(\ref{S_Q_simple_red2}). In fact, eqs.(\ref{c12}) mean the following relation
\begin{eqnarray}
C^{(12)} R^{(2)} +L^{(2)}(C^{(21)}+C^{(22)} R^{(2)})+I_2=0
\end{eqnarray}

Since all $p_i$ are positive, solution $V(t)$ (\ref{V11_1}) has no singularities  unless 
\newline $\sum_{m_1,m_2=1}^2 |t_{m_1m_2}|\to \infty$. However, 
offdiagonal elements of $V$ tend to infinity in some directions  in the space of parameters $t_{m_1m_2}$. Thus, $V$ is not bounded solution. 
Now we derive a simple example of the bounded soliton-kink solution. For this purpose we 
 take
\begin{eqnarray}
s_3=s_5=0,\;\;s_4=-1
\end{eqnarray}
instead of eqs.(\ref{ss1}).
Then one gets the following formula for $V$:
\begin{eqnarray}\label{V11}
&&
V(t)=
\left[
\begin{array}{cc}\displaystyle
\frac{f_{11}}{d + 5 e^{\eta_{12}+\eta_{21}}} 
 & \displaystyle\frac{f_{12} }{d e^{-\eta_{12}}+5 e^{\eta_{21}}} \cr
\displaystyle\frac{f_{21} }{d e^{-\eta_{21}}+5 e^{\eta_{12}}} &
\displaystyle \frac{f_{22}}{d +5 e^{\eta_{12}+\eta_{21}}} 
\end{array}
\right],\\\nonumber
&&
f_{11}=\frac{6}{5}(3 s_1-2)(6 s_1+1)s_2 ,\;\;
f_{12}=-3 (6 s_1 +1)s_2,\\\nonumber
&&
f_{21}=-2 (3 s_1 -2)(3 s_2 +2),\;\;
f_{22}=6(3 s_1-2)(3 s_2+2)s_2 ,\\\nonumber
&&
d=-6(3 s_1-2)s_2,
\end{eqnarray}
where $\eta_{n_1n_2}$ are the linear functions of $t_{n_1n_2}$:
\begin{eqnarray}
&&
\eta_{12}=\sum_{n_1,n_2=1}^2 a_{n_1n_2}t_{n_1n_2},\;\;
\eta_{21}=\sum_{n_1,n_2=1}^2 b_{n_1n_2}t_{n_1n_2},
\\\nonumber
&&
a_{11}=\frac{6}{5}(s_1+1),\;\;
a_{12}=-\frac{1}{r_1r_2}(r_1(2+(3 s_2+2) l_2(c^{(21)}_2+c^{(22)}_2 r_2))+r_2(a_{11}-2))
,\\\nonumber
&&
a_{21}=\frac{c^{(21)}_{1}}{5} (6s_1+1)+\frac{1}{l_1}-\frac{2}{l_2} ,\;\;
a_{22}=\frac{1}{l_2r_1r_2}(r_1(2+(3 s_2+2) l_2(c^{(21)}_2+c^{(22)}_2 r_2))-r_2(a_{21}l_2+2)),
\\\nonumber
&&
b_{11}=3s_2+1,\;\;
b_{12}=-\frac{1}{5 r_1r_2}(r_2(5+(6 s_1+1) l_1(c^{(21)}_1+c^{(22)}_1 r_1))+5r_1(b_{11}-1))
,\\\nonumber
&&
b_{21}=c^{(21)}_{2} (3s_2+2)-\frac{1}{l_1}+\frac{2}{l_2},\;\;
b_{22}=\frac{1}{5 l_1r_1r_2}(r_2(5+(6 s_1+1) l_1(c^{(21)}_1+c^{(22)}_1 r_1))-5r_1(b_{21}l_1+1)).
\end{eqnarray}
If $d>0$, then the diagonal elements of matrix (\ref{V11}) are kinks, while offdiagonal elements tend to infinity in some directions in the space of parameters $t_{m_1m_2}$.
 In order to obtain bounded solution we require
\begin{eqnarray}\label{etaeta}
\eta_{12}=a\eta_{21}\;\;\Rightarrow \;\; a_{ij} =a b_{ij},\;\;a>0.
\end{eqnarray}
For the sake of simplicity, we solve eqs.(\ref{etaeta}) for the particular choice of the arbitrary parameters:
\begin{eqnarray}
r_1=2,\;\;r_2=3,\;\;l_1=4,\;\;l_2=5,\;\;c^{(21)}_1=6.
\end{eqnarray}
One has
\begin{eqnarray}
s_1&=& \frac{741 + 366  c^{(22)}_2 - 12  c^{(22)}_1 (7 + 24  c^{(22)}_2))}{
   8 (-559 + 108  c^{(22)}_2 + 36  c^{(22)}_1 (2 +  c^{(22)}_2))}, \\
 s_2&=&\Big(-(14521 - 181090  c^{(22)}_2 + 43200  (c^{(22)}_2)^2 + 
     144  (c^{(22)}_1)^2 (-1 + 20  c^{(22)}_2) + \\\nonumber
&&
40  c^{(22)}_1 (-20 + 9  c^{(22)}_2 + 
       360  (c^{(22)}_2)^2)\Big)\times\\\nonumber
&&
\Big(40 (-1 + 15  c^{(22)}_2) (-559 + 108  c^{(22)}_2 + 
     36  c^{(22)}_1 (2 +  c^{(22)}_2)\Big)^{-1}, \\
 a &=& \frac{246 - 3690  c^{(22)}_2}{-233 + 2160  c^{(22)}_2 + 
    36  c^{(22)}_1 (-1 + 20  c^{(22)}_2)},\\\label{cc21}
 c^{(21)}_2 &=& \frac{25 - 36 (3 +  c^{(22)}_1)  c^{(22)}_2}{
   -89 + 12  c^{(22)}_1}
\end{eqnarray}
In additin, one has to provide positivity of $a=p_1>0$ in eq.(\ref{V11}) and positivity of $d=p_2>0$.
This requirement yields constraint for $c^{(22)}_i$, $i=1,2$:
\begin{eqnarray}\label{c22}
c^{(22)}_1 &=& \big(1240  p_1 -  p_1^3 (725 + 144  p_2) -  p_1^2 (525 + 248  p_2) \pm\\\nonumber
&&
      p_1 (1 +  p_1) \sqrt{25 (-2 +  p_1)^2 - 20  p_1  p_2}\Big)\Big(
    48  p_1^2 (-5 +  p_1 (5 +  p_2))\Big)^{-1}, \\\nonumber
 c^{(22)}_2 &=& \Big(-1200  p_1 + 5  p_1^2 (-263 + 48  p_2) + 
     5  p_1^3 (497 + 100  p_2) \pm 
\\\nonumber
&&
3  p_1 (1 +  p_1) 
      \sqrt{25 (-2 +  p_1)^2 - 20  p_1  p_2}\Big)\times\\\nonumber
&&
\Big(
    60 (10  p_1 - 5  p_1^3 -  p_1^2 (15 + 2  p_2) \pm
       p_1 (1 +  p_1) \sqrt{25 (-2 +  p_1)^2 - 20  p_1  p_2})\Big)^{-1}.
\end{eqnarray}
In particular, if $p_1=1/2$ and $p_2=1$, one has
\begin{eqnarray}
&&
s_1=\frac{1\pm\sqrt{185}}{24},\;\;s_2=\frac{15\pm\sqrt{185}}{30},\;\;c^{(21)}_2=\frac{3(-145\pm 123\sqrt{185})}{2080}
,\\\nonumber
&&
c^{(22)}_1 = \frac{-2545 \mp 3 \sqrt{185}}{192}, \;\;
c^{(22)}_2 = \frac{70 \mp 93 \sqrt{185}}{780}
\end{eqnarray}
Now expression for $V$, eq.(\ref{V11}) reads
\begin{eqnarray}\label{V11_expl}
&&
V(t)=-
\left[
\begin{array}{cc}\displaystyle
\frac{5\pm\sqrt{185}}{20(1 + 5 e^{(a+1)\eta_{21}})} 
 & \displaystyle\frac{(13\pm\sqrt{185}) }{2( e^{-a\eta_{21}}+5 e^{\eta_{21}})} \cr
\displaystyle\frac{(-17\pm\sqrt{185}) }{2( e^{-\eta_{21}}+5 e^{a\eta_{21}})} &
\displaystyle \frac{(35 \pm \sqrt{185})}{10(1 +5 e^{(a+1)\eta_{21}})} 
\end{array}
\right],\;\;\eta_{21}=\sum_{n_1,n_2=1}^2 b_{n_1n_2} t_{n_1n_2}.
\end{eqnarray}
Appropriate expressions for $b_{n_1n_2}$ are following
\begin{eqnarray}
&&
b_{11}=\frac{25\pm\sqrt{185}}{10},\;\;
b_{12}=\frac{295\pm 61\sqrt{185}}{30},\\\nonumber
&&
b_{21}=\frac{3 \left(9\pm 2\sqrt{185}\right)}{10},\;\;
b_{22}=\frac{- 214\mp 43\sqrt{185}}{60}.
\end{eqnarray}
We see that diagonal elements of $V$ represent kinks, while off-diagonal elements represent non-symmetrically shaped solitons.

Note that solution (\ref{V11_expl}) causes some restrictions on the  coefficients of eq.(\ref{S_Q_simple_red2}), which are eqs.(\ref{cc21}) and (\ref{c22})

\section{Conclusions}
\label{Section:Conclusions}
We represent a new algorithm allowing one to construct a rich variety of particular solutions to a new class of nonlinear PDEs of any order in any dimensions. These equations can be considered as multidimensional generalizations of well known   $C$- and $S$-integrable equations.  We show that the solution space may be rich enough to provide complete integrability of some of these equations. However, the problem of complete integrability  requires futher study. 

The suggested algorithm allows evident generalizations. For instance, let us generalize constraint (\ref{constrain1}) as follows:
\begin{eqnarray}\label{constrain1gen}
\sum_{m,n=1}^{D}A^{(m)} *\tilde A^{(n)} B^{(mnp)} =\sum_{m=1}^{D}\tilde A^{(m)} P^{(mp)},\;\;p=1,\dots,D,\;\;
j=1,\dots,m_0,
\end{eqnarray}
where $B^{(mnp)}$ and $P^{(mn)}$ are some constant $n_0\times n_0$ matrices.
Then one can show that  $n_0\times n_0$ matrix functions $V^{(m)}(t)$ are solutions to the following system of nonlinear PDEs:
\begin{eqnarray}\label{Qgen}
&&
\sum_{m,n=1}^{D}\left[\left(
V^{(n)}_{t_m} +\left(V^{(m)}V^{(n)}+V^{(m)} {\cal{A}}^{(n)}\right)\right)
B^{(mnp)}\right]+ \sum_{m=1}^{D} V^{(m)} P^{(mp)} =0,\\\nonumber
&&p=1,\dots,D.
\end{eqnarray}
 
Remark that the physical application of some of the derived  PDEs is obvious. For instance, the multidimensional $N$-wave equation (\ref{S_Q_simple_red_intr}) appears in multiple-scale analisis of any physical  dispersion system. However, physical applications must  be considered in more details.

The author thanks Professor P.M.Santini for usefull discussions. This work is supported by the RFBR grants  10-01-00787 and 09-01-92439 and by the grant NS-4887.2008.2.


\end{document}